\pdfoutput=1

\documentclass[apj]{emulateapj}

\usepackage[bookmarks=true,colorlinks=true,citecolor=blue,linkcolor=magenta,urlcolor=cyan]{hyperref}

\usepackage{natbib}
\usepackage{amsmath}
\usepackage{amssymb}
\usepackage{subfigure}
\usepackage{url}
\usepackage{CJK}

\newcommand{\rmd}{ {\ \mathrm d} }
\renewcommand{\vec}[1]{ {\mathbf #1} }

\newcommand{\grad}{ {\bf \nabla } }

\newcommand{\Fig}{{Figure}}
\newcommand{\Figs}{{Figures}}

\newcommand{\divB}{\nabla\cdot\mathbf{B}}
\newcommand{\crlB}{\nabla\times\mathbf{B}}

\newcommand{\SDO}{{\it SDO}}

\graphicspath{{figs/}}

\slugcomment{Accepted for publication in ApJ}
\shorttitle{Formation and Eruption of an AR Sigmoid}
\shortauthors{Jiang et al.}

\begin{document}
\begin{CJK*}{UTF8}{gbsn}

\title{Formation and Eruption of an Active Region Sigmoid: I. Study by
  Nonlinear Force-Free Field Modeling}

\author{
  Chaowei Jiang \altaffilmark{1},
  S.~T. Wu \altaffilmark{2},
  Xueshang Feng \altaffilmark{1},
  Qiang Hu \altaffilmark{2}}
\email{cwjiang@spaceweather.ac.cn, fengx@spaceweather.ac.cn}

\altaffiltext{1}{SIGMA Weather Group, State Key Laboratory for Space
  Weather, Center for Space Science and Applied Research, Chinese
  Academy of Sciences, Beijing 100190}
\email{wus@uah.edu}

\altaffiltext{2}{Center for Space Plasma and Aeronomic Research, The
  University of Alabama in Huntsville, Huntsville, AL 35899, USA}

\begin{abstract}
  We present a magnetic analysis of the formation and eruption of an
  active region (AR) sigmoid in AR 11283 from 2011 September 4 to
  6. To follow the quasi-static evolution of the coronal magnetic
  field, we reconstruct a time sequence of static fields using a
  recently developed nonlinear force-free field model constrained by
  the \SDO/HMI vector magnetograms. A detailed analysis of the fields
  compared with the \SDO/AIA observations suggests the following
  scenario for the evolution of the region. Initially, a new bipole
  emerges into the negative polarity of a pre-existing bipolar AR,
  forming a null point topology between the two flux systems. A weakly
  twisted flux rope (FR) is then built up slowly in the embedded core
  region, largely through flux-cancellation photospheric
  reconnections, forming a bald patch separatrix surface (BPSS)
  separating the FR from its ambient field. The FR grows gradually
  until its axis runs into a torus instability (TI) domain near the
  end of the third day, and the BPSS also develops a fully
  S-shape. Unlike in the case of standard TI, the FR does not erupt
  instantly since it is still attached at the photosphere along the
  bald patch (BP) portion of the polarity inversion line. The combined
  effects of the TI-driven expansion of the FR and the line-tying at
  the BP tear the FR into two parts with the upper portion freely
  expelled and the lower portion remaining behind the post-flare
  arcades. This process dynamically perturbs the BPSS and results in
  the transient enhanced brightening of the sigmoid. The accelerated
  expansion of the upper portion of the FR strongly pushes its
  envelope flux near the null point and triggers breakout reconnection
  at the null, as evidenced by a remarkable circular flare ribbon,
  which further facilitates the eruption. We discuss the important
  implications of these results for the formation and disruption of
  sigmoid region with FR.
\end{abstract}


\section{Introduction}
\label{sec:intro}

Sigmoid, the name given to forward or inverse S-shaped coronal loops
seen often in soft X-ray (SXR) and sometimes in extreme ultraviolet
(EUV) emission \citep{Rust1996, Gibson2002}, is one of the most
important precursor structures for solar eruptions \citep{Hudson1998,
  Canfield1999, Canfield2007}. Thus, it is prudent to concentrate on
sigmoidal regions if one wants to understand the structure and
evolution of erupting regions, due to their higher probability for
producing eruptions. This is not surprising since the shape of
sigmoids usually indicates sheared and twisted magnetic structures,
which carry a field-aligned current and thus free magnetic
energy. Most sigmoids appear in active regions (ARs), usually situated
on top of a curved polarity inversion line (PIL). They have been
envisioned as the twisted and sheared core field, e.g., magnetic flux
rope (FR), embedded in a potential envelope field \citep{Moore1992},
which stabilizes the core field against eruption. In morphology, four
kinds of sigmoids can be distinguished, i.e., multiple loops,
inter-region loops, single S-shaped and double J-shaped loops
\citep{Pevtsov2002}. Also they can be collectively referred to as two
types, transient and long-lived sigmoids, respectively. Transient
sigmoids are sharp and bright, and usually become clearly noticeable
only for a short time before the actual eruption, while long-lived
sigmoids appear more diffuse and can survive for many hours or even
days until the eventual eruption \citep{Green2007,
  McKenzie2008}. Usually sigmoids evolve into cusps and post-flare
arcade of loops during eruption, but in some cases they survive and
remain even after eruption \citep{Pevtsov2002,Gibson2002}.

Sigmoids are considered to result from enhanced current dissipation
that accumulates hot plasma along correspondingly shaped field
lines. Due to the extremely low resistivity of the coronal plasma,
these currents need to take the form of thin layers for the
resistivity to be important for sufficient dissipation. Narrow current
sheets can easily form along magnetic interface layers, e.g., magnetic
separatrices and quasi-separatrix layers (QSLs), across which the
connectivity of field lines discontinues or changes abruptly \citep{Demoulin2006,Demoulin2007}. The
interfaces separating a coronal FR from its ambient field usually form
the sigmoidal shape when observed from above, and they are naturally
invoked in different models of sigmoids \citep[e.g., see review
of][]{Gibson2006}. In particular, with an analytical force-free FR
model embedded in an arcade field, \citet{Titov1999} showed that in
the process of the FR emerging rigidly into the corona, there is a
separatrix surface touching the photosphere along sections of the PIL,
where the transverse magnetic fields cross from the negative to the
positive polarity (opposite to a potential field case).  These
sections of the PIL are called bald patches (BPs), and the BP
separatrix surface (BPSS) has an S-shape viewed from above,
which speaks to its potential importance in producing a sigmoid shape.
Even in the later
phase of emergence, the S-shaped BPSS bifurcates into a double J-shaped
QSL with the main body of the FR elevated off the
photosphere, the sigmoidal shape remains, which matches the QSL
\citep{Aulanier2010}.

Investigations based on numerical simulations have shown that
generally a FR does not emerge bodily from below the photosphere, but
forms in situ in the corona \citep{Magara2006, Fan2009}.
Nevertheless, those FR-related BPSS and QSL mentioned above are still
essential to the FR formation and eruption processes, since they
relies on magnetic reconnections occurring in these thin layers. It
has been shown that flux convergence and cancellation yield the
formation of a FR through tether-cutting-like reconnection occurring
on the photosphere between previously sheared arcades
\citep{Ballegooijen1989}. This process forms BPs in the PIL, and the
sheared field lines reconnect at the BP, pass through the BPSS and
form S-shaped field lines in the rope. With the increase of magnetic
pressure in the growing FR, its main body will bulge upward and might
detach itself from the photosphere. In such a case, the BPSS
bifurcates gradually and transforms into the double J-shaped QSL, the
cross section of which contains an X-line-like configuration in the
corona, also known as the hyperbolic flux tube \citep[HFT;
see][]{Titov2002}. Now the standard tether cutting reconnection
\citep{Moore1992, Moore2001, Janvier2013, LiuC2013} settles in, which
occurs in the corona at the HFT, as another important mechanism for
further building up of the FR from the sheared arcades.

The question of how the eruption of a sigmoidal FR is initiated and
driven is still under intense debate \citep[see, e.g.,][and references
therein]{Aulanier2010, Schmieder2013, Aulanier2013}. In addition to
their roles played in the FR formation, both the flux cancellation and
tether cutting have also been invoked widely to account for triggering
the loss of equilibrium of the FR by increasing the magnetic pressure
in the rope and, in turn, reducing the restraining tension force of
the envelope field \citep{Ballegooijen1989, Moore2001, Amari2003a,
  Amari2003b, Linker2003}. However, with a detailed
magnetohydrodynamic (MHD) simulation of an eruption initiated in a
sheared bipolar region with photospheric field diffusion (emulating
the flux cancellation), \citet{Aulanier2010} found that neither of
these processes triggers an eruption, but actually, the eruption is
caused by a kind of ideal MHD instability of the FR-overlying flux
system, i.e., the torus instability \citep[TI;
see][]{Kliem2006}. Generally, a coronal FR can be regarded as an
anchored partial current ring, which experiences an outward ``hoop
force'' due to the self-repelling force of the image current under the
photosphere. This hoop force is counteracted by external potential or
sheared field, i.e., the overlaying arcade. The TI describes the
instability due to expansion of the FR if the restoring force of the
external field decreases with the height faster than the hoop
force. This instability can be characterized by a decay index of the
external field with an unstable threshold of $\sim 1.5$
\citep{Torok2007,Aulanier2010}

Several other models besides the TI have been proposed for the
initiation mechanism of the FR eruptions. The kink instability (KI)
for a FR occurs if the twist, a measure of the number of windings of
the field lines about the rope axis, exceeds a critical value (about
1.5--2), leading to a helical deformation of the FR's axis
\citep{Hood1981, Velli1990, Torok2004, Torok2005}. Numerical models
have shown that the KI can generate sigmoidal current sheet just prior
to an eruption at the interface of the FR and its ambient field
\citep{Kliem2004, Fan2004}, and thus suggests another explanation for
transient sigmoids.

The breakout model \citep{Antiochos1999} provides another possibility
for eruptions by directly removing the overlying constraining flux
through reconnection above the core field. In the original breakout
model based on a quadrupolar magnetic field configuration, the
eruption is triggered by reconnection at a coronal null above the
sheared core arcade, which removes the overlying flux and then allows
the core to escape, similar to the streamer and flux model
\citep{Wu1997}. There are also lateral versions of breakout models
proposed \citep[e.g.,][]{Chen2000, Lin2001}, in which the coronal null
point reconnection takes place on the side of the eruptive core
instead of above it. More complex breakout models have been
investigated \citep[e.g.,][]{Roussev2007A, Jacobs2009} which included
much more realistic magnetic topology with multiple null points. The
critical building block of these models is that the overlying flux
contains coronal null point, which is possibly formed by a new flux
emerging into an inverse preexisting field \citep{Wu2005,
  Torok2009}. It should be noted that although the breakout model
generally does not involve a pre-eruption FR and thus sigmoid, the
breakout reconnection should work for either a sheared arcade or a FR
in the core field, as long as there is null point existing in the
overlying flux.

The aim of this paper is to provide a comprehensive study of the whole
process of the formation and eruption of a sigmoidal FR, and to
identify the different mechanisms for the FR built-up and
eruption. Our study is based on observations and a nonlinear
force-free field (NLFFF) model. Although observations with increasing
resolutions have shed important light on this process
\citep[e.g.,][]{Schmieder2012, Zhang2012, Cheng2013}, the critical
parameter, i.e., the coronal field, is difficult to measure
at least at present. NLFFF extrapolation has been accepted as a viable
tool to obtain the pre-eruption, near-static coronal field due to the
low-$\beta$ (ratio of plasma pressure to magnetic pressure) nature of
the corona \citep{Schrijver2008, Wiegelmann2008, Su2011}. Based on a
time-sequence of static force-free fields, one can further study the
slow build-up process of the field prior to eruption by assuming that
such evolution can be described by successive equilibria
\citep{Regnier2006, Wu2009, Jiang2012c, Sun2012}. This method is
justified by the fact that the evolution of the coronal field, driven
by the photospheric motion with a flow speed less than several
km~s$^{-1}$, is sufficiently slow compared with the speed for the
coronal field relaxing to equilibrium (i.e., the coronal Alfv\'en speed),
which is up to thousands of km~s$^{-1}$ \citep{Antiochos1987,
  Seehafer1994}. Although the dynamic evolution of the field
intrinsically requires an MHD simulation driven by photospheric fields
and flows, such a data-driven simulation \citep{Wu2004,Wu2006} is
generally prohibitive in this type of study due to the very long time
of slow evolution and very high Alfv\'en speed in the corona, which is
difficult to handle in numerical schemes \citep{Jiang2012c}. Thus it
is more practical to use the successive extrapolated fields to compare
with an idealized MHD simulation for our purpose, as was done by
\citet{Savcheva2012b}.

The studied event is the formation of a sigmoidal FR in AR 11283 from
2011 September 4 to 6 and its eruption. As shown in the following,
this event involves a variety of magnetic processes including flux
emergence, flux cancellation, ideal MHD instability and breakout
reconnection.  It thus provides an attractive sample for our study. We
present a NLFFF model to study the slow evolution of the magnetic
field of the region over three days leading to an X-class flare on
2011 September 6. Our attention is focused on the building up process
of the magnetic topology and non-potentiality of the pre-eruption
field and the initiation mechanism of the eruption. In Section~2,
relevant observations of AR 11283 during the time of our interest is
described briefly. We then analyze the magnetic evolution prior to and
through the sigmoid eruption, in Section~3 for the photospheric field
and Section~4 for the coronal field, based on the \SDO/HMI
magnetograms and our NLFFF model. In Section~5 we study the initiation
mechanism of the eruption. Section~6 is devoted to the analysis of the
evolution of magnetic energy. The dynamic evolution of the field
during eruption cannot be reproduced by the static extrapolation and
strongly requires an MHD simulation
\citep{Jiang2013apjl,Kliem2013}\footnote{A preliminary MHD simulation
  of this eruption process has been carried out by
  \citet{Jiang2013apjl}}. In the second paper of this series, we will
conduct a full MHD simulation and a detailed analysis of the fast
magnetic evolution during the sigmoid eruption.

\begin{figure*}[htbp]
  \centering
  \includegraphics[width=0.8\textwidth]{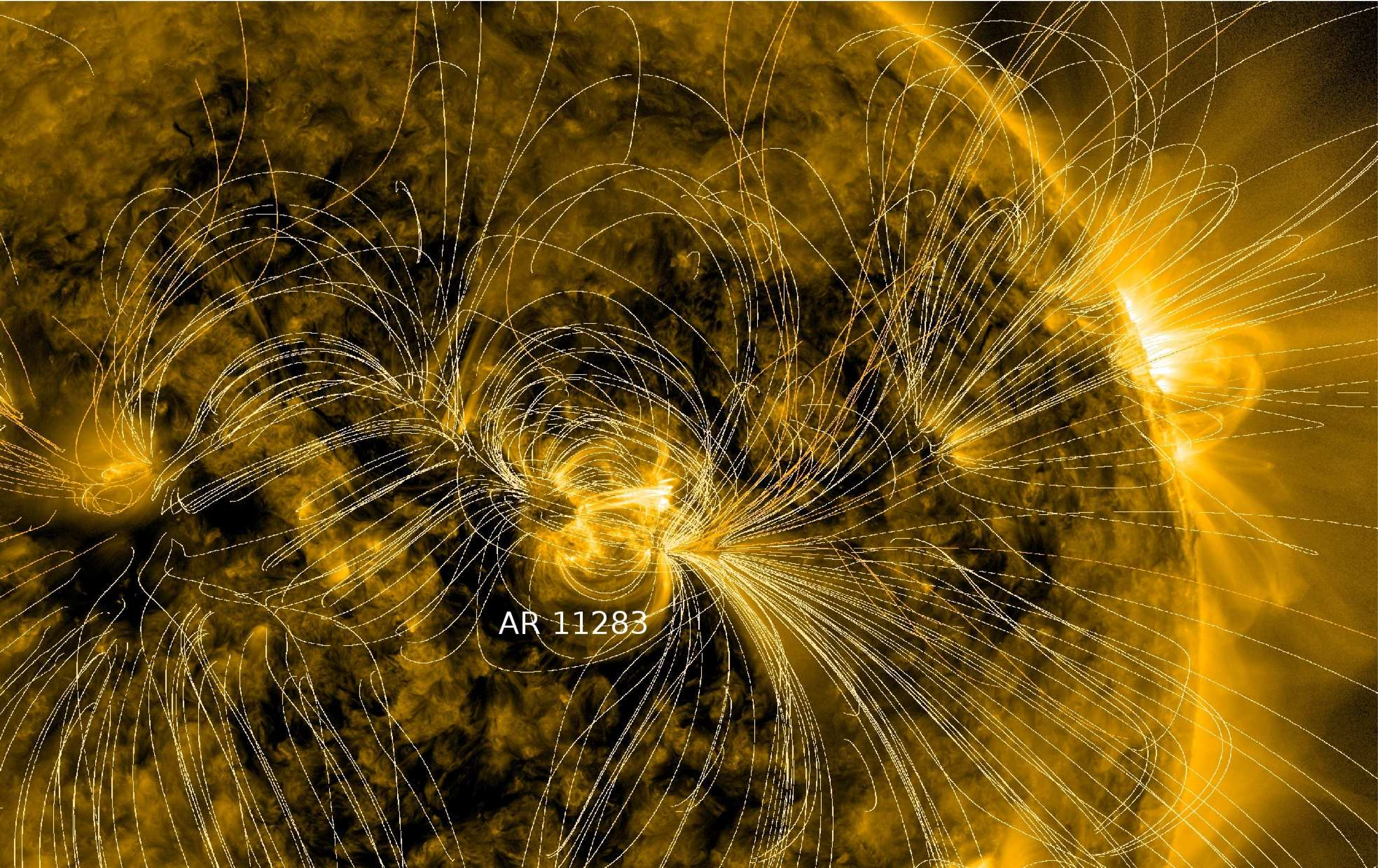}
  \caption{Part of a full-disk \SDO/AIA-171 image taken on the end of
    2011 September 6. Overlying is the PFSS (potential field source
    surface) field lines to show the large-scale magnetic environment
    of AR~11283.}
  \label{fig:1.0}
\end{figure*}

\begin{figure}[htbp]
  \centering
  \includegraphics[width=0.48\textwidth]{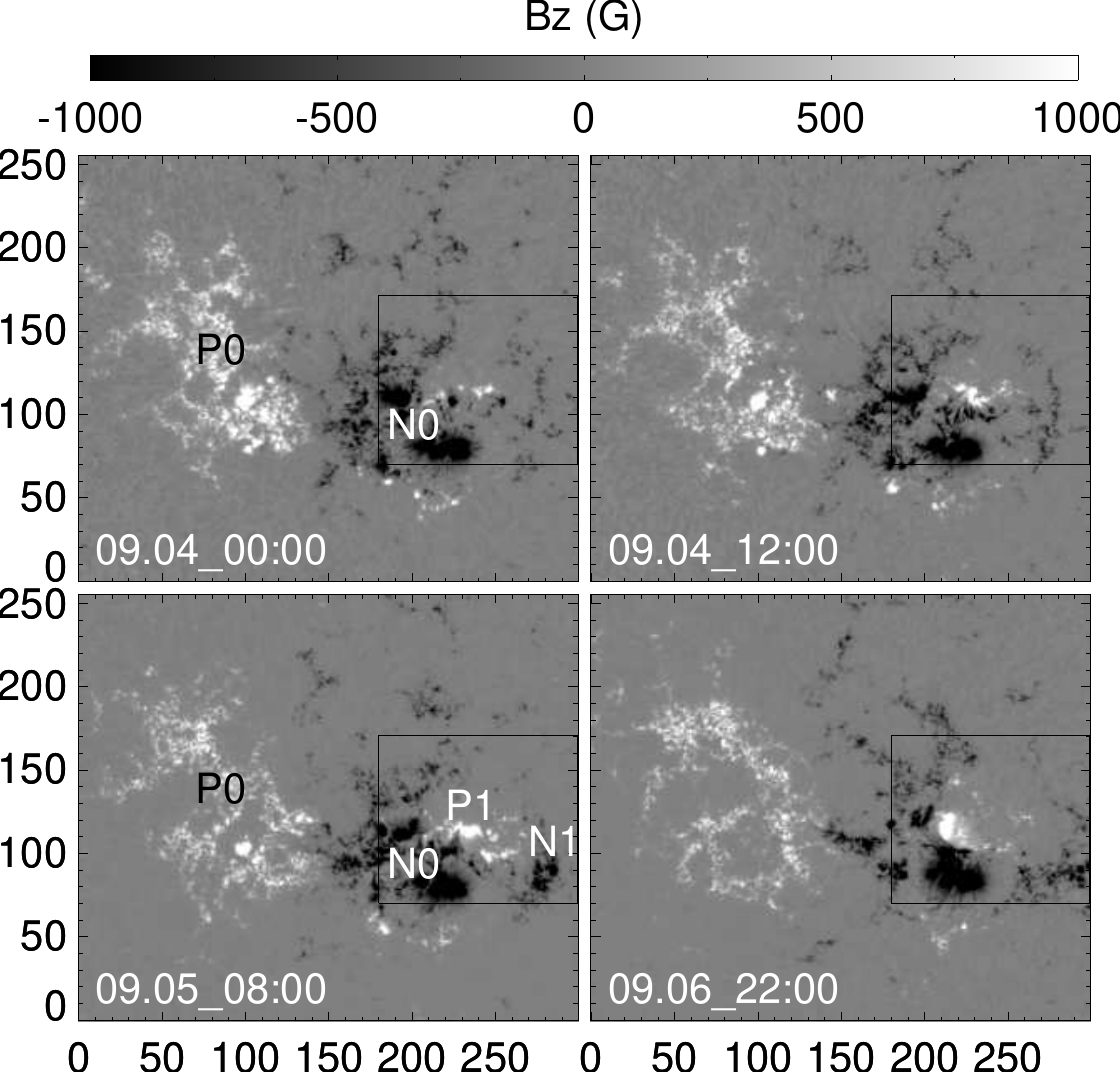}
  \caption{Evolution of the photospheric field. Four snapshots are
    shown and {\it an animation is available}. The black boxes outline
    the region of the flux emergence and eruption. P0/N0 label the
    pre-existing polarities and P1/N1 the new emerging
    polarities. The length unit is arcsec.}
  \label{magram_evolve}
\end{figure}

\begin{figure*}[htbp]
  \centering
  \includegraphics[width=\textwidth]{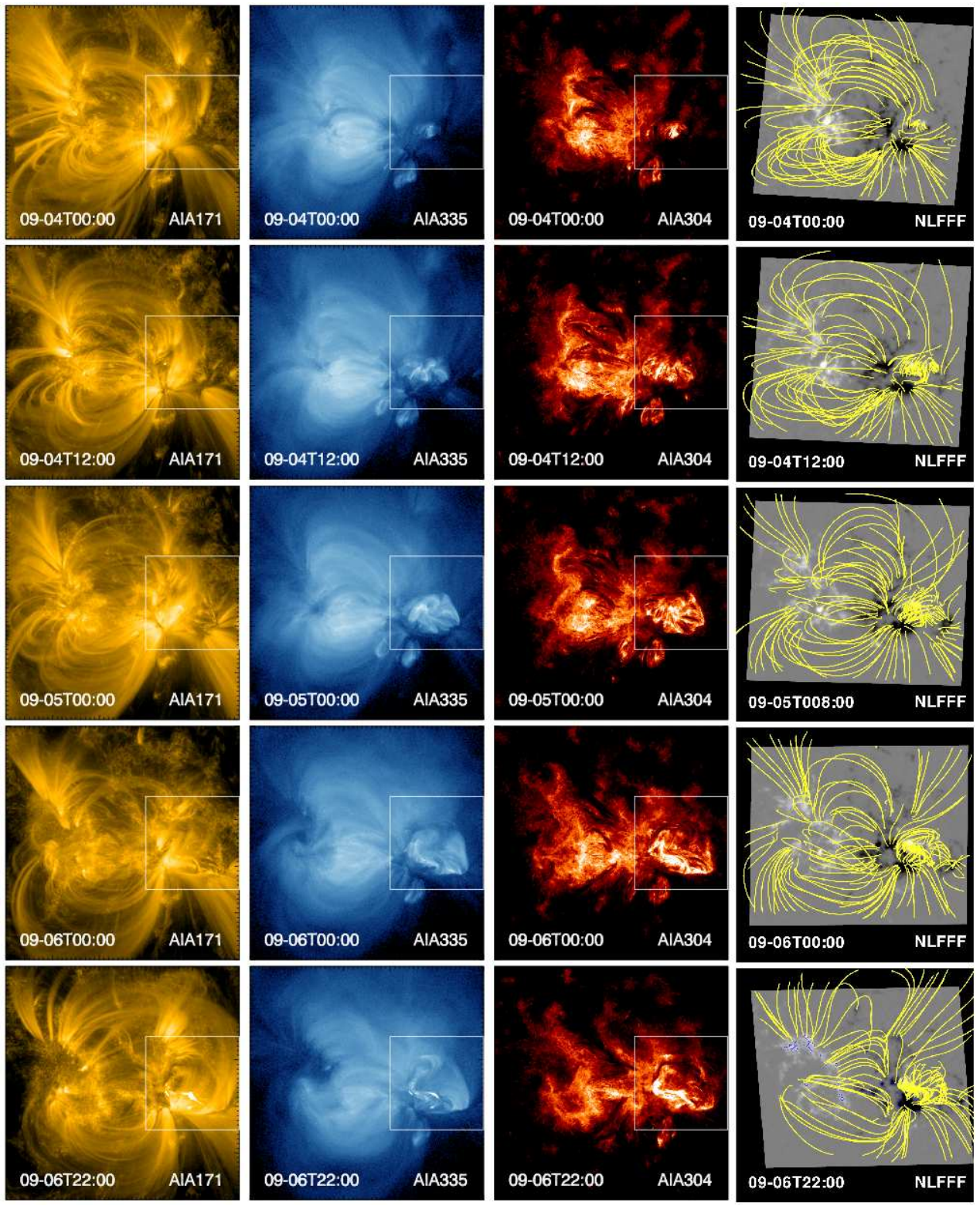}
  \caption{Evolution of the AR observed in different AIA channels and
    modeled by NLFFF extrapolations (yellow lines are the magnetic
    field lines and $B_z$ is shown on the bottom).  The boxes outline
    the flux emergence site. View angles of the NLFFFs are aligned
    with the AIA images.}
  \label{fig:AIA+NLFFF_evolve}
\end{figure*}

\begin{figure}[htbp]
  \centering
  \includegraphics[width=0.45\textwidth]{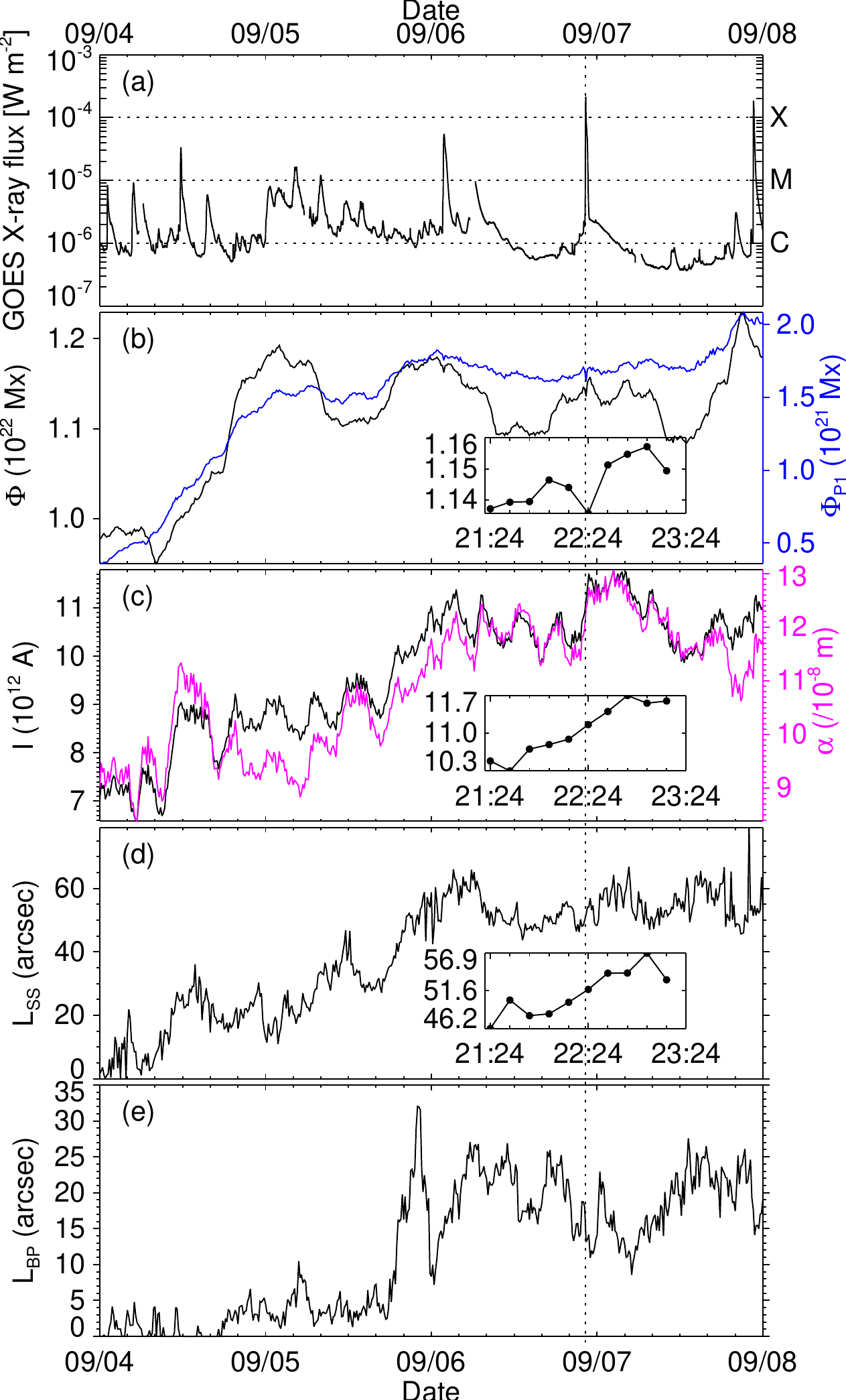}
  \caption{Evolution of different parameters over 4 days.  (a) {\it
      GOES} soft-X ray flux (1--8~{\AA} channel). (b) Total unsigned
    magnetic flux $\Phi$ and the new emerging positive flux $\Phi_{\rm
      P1}$. (c) Total unsigned current $I$ and the torsion parameter
    $\alpha$. (d) Length of strong sheared PIL: $L_{\rm SS}$. (e)
    Length of BP on the main PIL: $L_{\rm BP}$. The vertical dashed
    lines indicate the peak time of the X2.1 flare at 22:20~UT on
    September 6. Inserts of (b), (d), and (d) show results with 2~h
    around the flare peak time.}
  \label{fig:linevol}
\end{figure}

\section{Observations}
\label{sec:obs}

AR 11283 is one of the very productive ARs in the new solar cycle. It
has produced several major flares/CMEs when near the disk center from
2011 September 5 to 7. \Fig~\ref{fig:1.0} shows the AR's location on
the disk on September 6. Potential field lines overlaid on the AIA-171
image illustrate the large-scale magnetic environment of the AR. We
study the evolution of this region from September 4 to 6.
\Fig~\ref{magram_evolve} gives the evolution of the photospheric
magnetic flux distribution observed by \SDO/HMI. Initially, on
September 4, this region has a simple bipolar configuration as a
mature AR, with a leading negative-polarity sunspot N0 and a following
positive polarity P0 that appears much more dispersed. Evolution of
the photospheric field is then dominated by a new bipole (labeled as
P1/N1) emerging into the west of the preexisting negative polarity N0,
forming a delta sunspot group. The new polarities move apart from
each other quickly, as usually observed in flux emergence site,
and N1 progressively approaches the west boundary of the
magnetogram. Note that the positive P1 is surrounded by negative flux,
indicating that a coronal null point is likely formed above. After the
initial stage of flux emergence with the apparent new flux injection
finished, rotation of P1 and shearing motion between P1/N0 are
observed. Such photospheric motions make a continuous injection of
magnetic free energy and helicity into the corona.  Meanwhile,
cancellation of flux elements with inverse polarities along the PIL
can be clearly seen by inspecting the animation of
\Fig~\ref{magram_evolve}.

\Fig~\ref{fig:AIA+NLFFF_evolve} shows evolution of the EUV emissions
observed by \SDO/AIA, which also reflects well the emergence
process. Roughly outlined by the enhanced emission, the basic magnetic
topology evolves from a single field-line connectivity domain of
P0--N0 (excluding some outer loops possibly connecting to
surrounding ARs) to two domains of different connectivity. The
new domain appears to be embedded in the pre-existing one.
The separatrix of the two connectivity domains manifests itself rather
clearly in the AIA-335 images (denoted by the boxes in
\Fig~\ref{fig:AIA+NLFFF_evolve}), showing roughly a closed circular
shape expanding with time. During the emergence process, small flares
and jet-like features are frequently observed, which is likely due to
reconnection accounting for the coronal field reconfiguration. The
flares also demonstrate circular ribbons outlining the magnetic
separatrix, which strongly implies the presence of a fan-spine
topology associated with a coronal null \citep{Masson2009, Wang2012,
  Jiang2013apjl, Sun2013}. The embedded core field is significantly
sheared on September 6, and by the end of the day, S-shaped loops
appear progressively from the highly sheared arcades above the PIL.

The most powerful eruption occurs in the embedded region near the end
of September 6. An X2.1 flare starts at 22:12~UT, peaks at 22:20~UT
and ends at 22:24~UT, as shown by the GOES flux in
\Fig~\ref{fig:linevol} (a). The flare ribbons consist of two different
components, an enhanced circular ribbon at the same location of the
magnetic separatrix around the new emerged core, and a standard flare
ribbon along the main PIL between P1/N0
\citep{Jiang2013apjl,Dai2013}. A remarkable S-shaped sigmoid brightens
a few minutes before the flare, followed by a drastic eruption of FR
and associated filament toward the northwest that evolves into a CME
\citep{FengL2013,Jiang2013apjl}. Significant movement of the filament
persists till the end of the day. The sigmoidal field reforms shortly
after this eruption, and another X1.8 flare occurs at the same site on
the following day, with a similar pre-flare sigmoid observed.  Our
study is focused on the magnetic field evolution leading to the
sigmoid eruption event on September 6.

In the following sections we analyze the long-term magnetic evolution
prior to the eruption, to show how the non-potentiality accumulates
and the sigmoidal field forms. Attention to the abrupt change of the
field across the major flare is also paid. The investigation is
performed first for the photospheric field and then the
three-dimensional (3D) coronal field.

\begin{figure*}[htbp]
  \centering
  \includegraphics[width=0.8\textwidth]{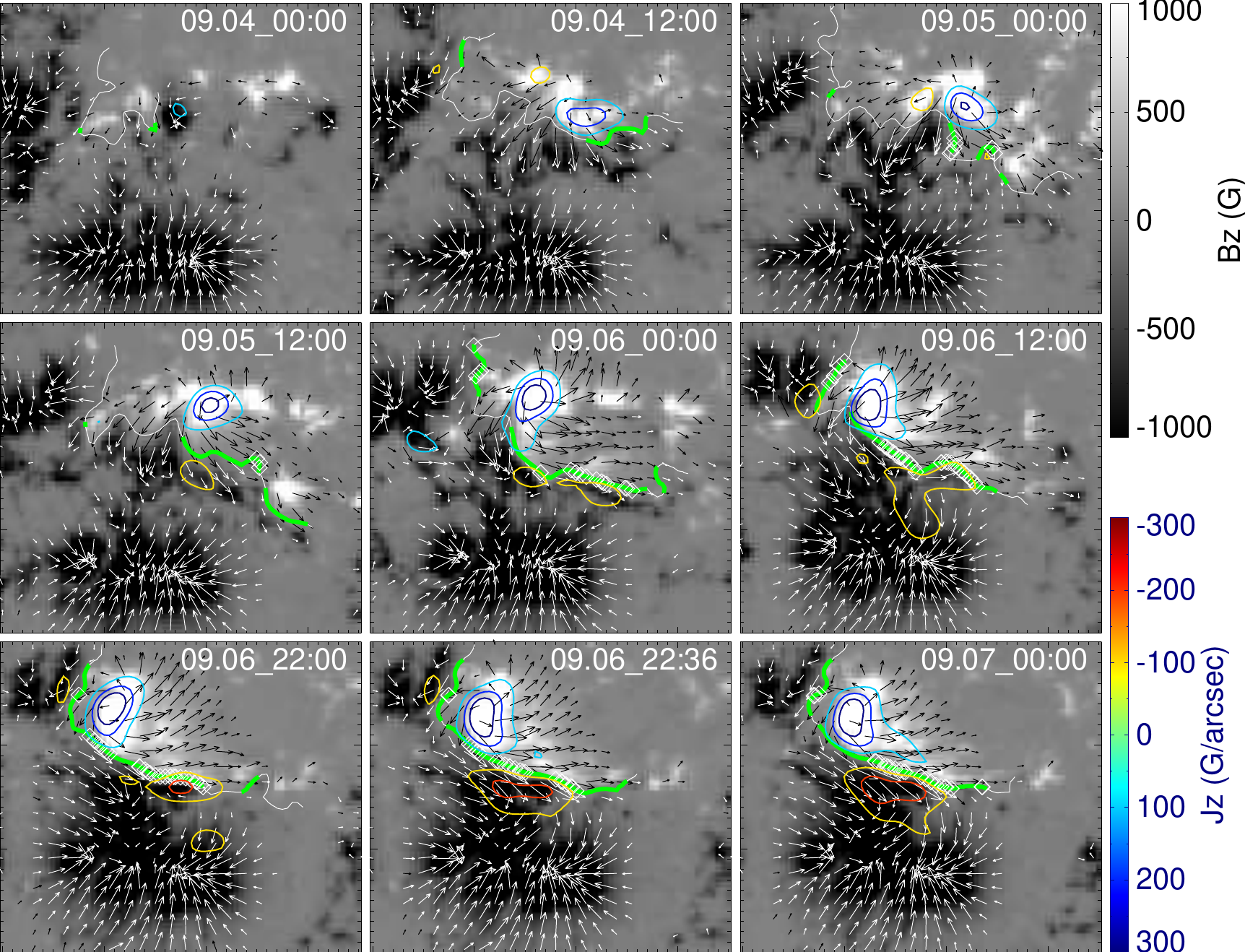}
  \caption{Evolution of the photospheric vector field in the core
    region. The vectors represent transverse field (above 200~G), with
    longest vector as 1700~G. Color closed contour lines are for the
    vertical current density $J_{z}$ derived from 5-pixel
    Gaussian-smoothed vector magnetogram. The white lines are the main
    PIL and the thick green segments denote the strong magnetic shear
    (shear angle $>45^{\circ}$) parts with strong transverse field ($>300$~G). The
    length of these parts of the main PIL is $L_{\rm SS}$. The white
    diamonds overlaid on the PIL represent the locations of BP.}
  \label{local_magram}
\end{figure*}


\section{Evolution of Photospheric Field}

Based on the vector magnetograms observed by \SDO/HMI \footnote{The data is
  downloaded from website
  \url{http://jsoc.stanford.edu/jsocwiki/ReleaseNotes2}, where
  products of HMI vector magnetic field datas are released for several
  ARs.} with a cadence of 12~min, we plot the evolution of a set of
parameters in \Fig~\ref{fig:linevol}, including the total unsigned
flux $\Phi = \int_{S}|B_{z}|\rmd s$ (where $S$ represents the
photosphere) and unsigned current $I = \int_{S}|j_{z}|\rmd s$, the
torsion parameter $\alpha=\mu I/\Phi$ and the length of strong sheared
PIL, $L_{\rm SS}$, which is defined as the portions of the main PIL
with shear angle $ > 45^{\circ}$ and transverse field $ > 300$~G
\citep{Falconer2002,Wu2009}. We focus on the field in the emergence/eruption region
outlined by the black box on the magnetograms in
\Fig~\ref{magram_evolve}. The current distributions and the locations
of strong sheared PIL are overlaid on the vector magnetograms as shown
in \Fig~\ref{local_magram}.

As shown in \Fig~\ref{fig:linevol} (b), the total magnetic flux
increases mainly on the first day and then decreases episodically and
slightly because of the combined effects of flux emergence and
cancellation. We also compute the flux of the new emerging P1,
$\Phi_{\rm P1}$, which is a more suitable monitor for the flux
emergence. $\Phi_{\rm P1}$ increases fast on the first day, climbing
from $\sim 0.3$ to $\sim 1.5\times 10^{21}$~Mx and shows no
significant increase afterward. Although interrupted repeatedly by
small flares, the non-potentiality parameters, i.e., the total current
$I$, field twist $\alpha$ and magnetic shear $L_{\rm SS}$ (panels (c)
and (d) of \Fig~\ref{fig:linevol}), show an evolution trend of
continuous increasing until September 6. While the flux emergence is
dominant in the first day, the most significant increase of the
non-potentiality occurs on the second day (September 5), during which
the photospheric shear and rotation are most clearly observed
(\Fig~\ref{local_magram}). This demonstrates that the successive
accumulation of non-potentiality is mainly due to the surface motion
(shear/twist) on the photosphere but not the flux
emergence.

We do not find a distinct decrease of
the photospheric non-potentiality parameters through the major flare (see the inserts in
panels (c) and (d) of \Fig~\ref{fig:linevol} for the two hours evolution
around the flare peak time). On the contrary they increase
slightly, which is possibly a result of the so-called `implosion' effect on the
photosphere by the eruption \citep{Hudson2000}, which enhances the
transverse field and holds the magnetic stress (associated with
photospheric magnetic shear) during the flare
\citep[e.g.,][]{WangS2012,Liu2012,Wang2013}. Inspection of \Fig~\ref{local_magram}
shows that the transverse field near the PIL indeed increases from
22:00~UT to 22:36~UT. A study of the photospheric abrupt change during
this flare has been carefully performed by \citet{Petrie2012} and the
same conclusion was drawn. As a consequence, it might be difficult to capture the features of releasing non-potentiality during the flare solely in the photospheric field, thus a point of view from the 3D coronal field is
required. The enhanced transverse field suggests a
shortening of the core field lines (possibly due to the flare related
reconnection). This is confirmed by the rapid change of the vertical
current distribution during the flare. As shown in
\Fig~\ref{local_magram}, prior to the eruption, a pair of reversed
strong current concentrations are located beside the main PIL and they
approach much closer after the flare. According to the near force-free
assumption that coronal currents follow along field lines, there
should be field lines (possibly a sheared FR) connecting the pair
of current concentrations, and the approaching of them naturally
indicates a shortening of the field lines connecting them.

Based on the vector magnetogram, we can find the BPs on the main PIL
where the transverse fields cross the PIL from negative to positive
flux.
Since the photosphere can be
regarded as a line-tying boundary for the coronal field, a BP field line
is thus very special because it is anchored at the BP in addition to its two
footpoints. As a result, the continuous set of BP field lines
forms a BPSS and divides the coronal field into different topological
domains. Thus finding the locations of BP is important to study
the magnetic topology of the coronal field. Furthermore, the BP is
usually a indicator of the presence of FR in the corona
\citep{Titov1999, Aulanier2010}, and thus is essential in the studied
event. We compute the BP locations according to the condition $\vec
B\cdot\grad B_{z}>0$ on the main PIL and plot them in \Fig~\ref{local_magram}. We
also plot the time evolution of the total length of the BPs, $L_{\rm BP}$,
in \Fig~\ref{fig:linevol}~(e). The plots show that extended BPs appear and
develop on the central portion of the main PIL after about 12:00~UT on
September 5, a half day later than the beginning of substantial
increasing of the strong shear length $L_{\rm SS}$. Within hours, the
BP length grows to more than half of the $L_{\rm SS}$. It shrinks
slightly through the eruption, but never disappears. Due to the
weak twisting of the related FR,
the BP is usually broken into several segments.
We study the BP evolution further in the following with relevant coronal
fields.

\section{Development of Coronal Field}

\subsection{Field Extrapolation Method}

In the duration without dynamic eruptions, the coronal field evolution
can be regarded as quasi-static, and can be modeled
well by the NLFFF model. The basic assumption of the NLFFF model is that
the Lorentz force is self-balancing in the corona because of its low-$\beta$
environment. To solve the general NLFFF problem
\begin{equation}
  \label{eq:ff}
  (\crlB) \times \vec B = \vec 0,\ \
  \divB = 0,
\end{equation}
we have developed an MHD-relaxation-based code, CESE--MHD--NLFFF
\citep{Jiang2013NLFFF}. We start from a potential field based on the
normal component of the magnetogram, then replace the field at the
bottom boundary with the vector magnetogram, and use a zero-$\beta$ MHD
relaxation technique to achieve the final force-free state. To improve
the convergence speed of the MHD relaxation, we employ an advanced
conservation-element/solution-element (CESE) space-time scheme
implemented on a block-structured, non-uniform adaptive grid with
parallel computation \citep{Jiang2010}. For a detailed description of
the CESE--MHD--NLFFF code please refer to \citep{Jiang2012apj,
  Jiang2013NLFFF}.

Since the photospheric field contains force (due to high-$\beta$) and
measurement noise, a new preprocessing code \citep{Jiang2013SoP} is
applied to remove the force and smooth the raw data, which provides a
more consistent magnetogram for the NLFFF model. In our extrapolation,
we first preprocess the raw magnetograms with their original
resolution (i.e., 0.5~arcsec), and then rebin them to 1~arcsec/pixel
as the final input for the extrapolation code. The field of view (FoV)
of the vector magnetograms is $300\times 256$~arcsec$^{2}$. To reduce
the numerical boundary influence, we use a larger extrapolation box of
$448\times 384\times 320$~arcsec$^{3}$ which includes a peripheral
region
of 64~arcsec width around the vector magnetogram, and the side and top
boundaries of this larger extrapolation box are fixed with the
potential field value during the relaxation process.

For the long-term evolution of four days, we extrapolate the fields
with a cadence of 4~h. For the 2~h around the X2.1 flare (from
21:24~UT to 23:24~UT on September 6) extrapolation is performed for
the full 12-min cadence. In \Fig~\ref{fig:AIA+NLFFF_evolve}, the 3D
field lines are plotted for five snapshots before the
eruption. Comparison with the AIA-171 loops shows that the field lines
resemble well the coronal loops in each time, demonstrating the
validation of using the NLFFF model to reproduce the slow evolution of
the coronal field. There are also some large loops not reconstructed
well, and these loops are closed connecting to the northwest of the AR, which is
out of the magnetogram's FoV (see the larger FoV image in
\Fig~\ref{fig:1.0}). Such misalignment between extrapolation and
observation results from the flux imbalance of the vector
magnetogram, which has an average value of $-10\%$
\citep{Jiang2013NLFFF}, and a spherical extrapolation with a larger
magnetogram would be better for these long-connecting loops
\citep{Jiang2012apj1, Savcheva2012a}. Nevertheless,
\citet{Jiang2013NLFFF} have demonstrated that the current code can
reconstruct the AR's core field excellently (e.g., the flux emergence site
here).

\begin{figure*}[htbp]
  \centering
  \includegraphics[width=0.8\textwidth]{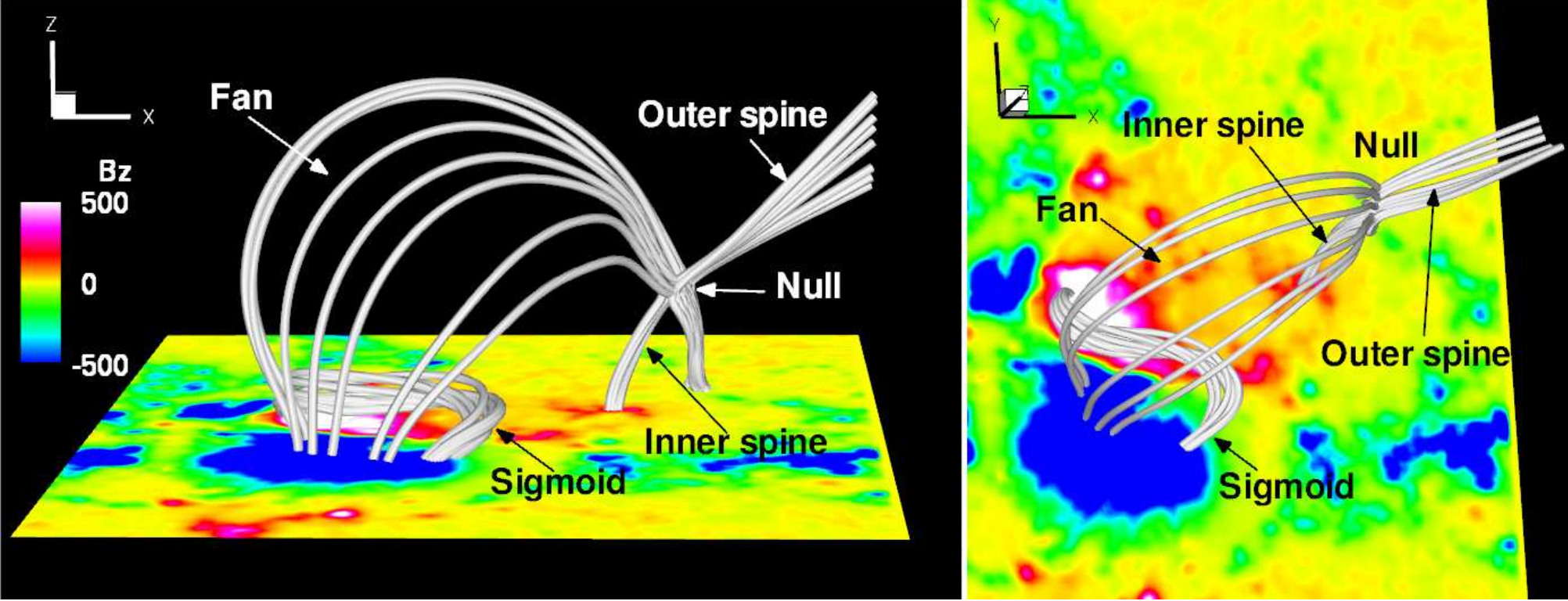}
  \caption{Magnetic topology based on NLFFF extrapolation for the
    pre-eruption field at 22:00~UT on September 6. The sigmoid field
    is the low-lying S-shaped lines; field lines closely touching the
    null outline the spine-fan topology of the null, at which the lines
    form a X-point configuration. The null point is situated at about
    18~arcsec (13~Mm) above the photosphere and 50~arcsec away
    from the sigmoid in the same direction of the eruption. The left
    panel is a side view and the right is the {\SDO} view.}
  \label{basic_topology}
\end{figure*}

\begin{figure*}[htbp]
  \centering
  \includegraphics[width=\textwidth]{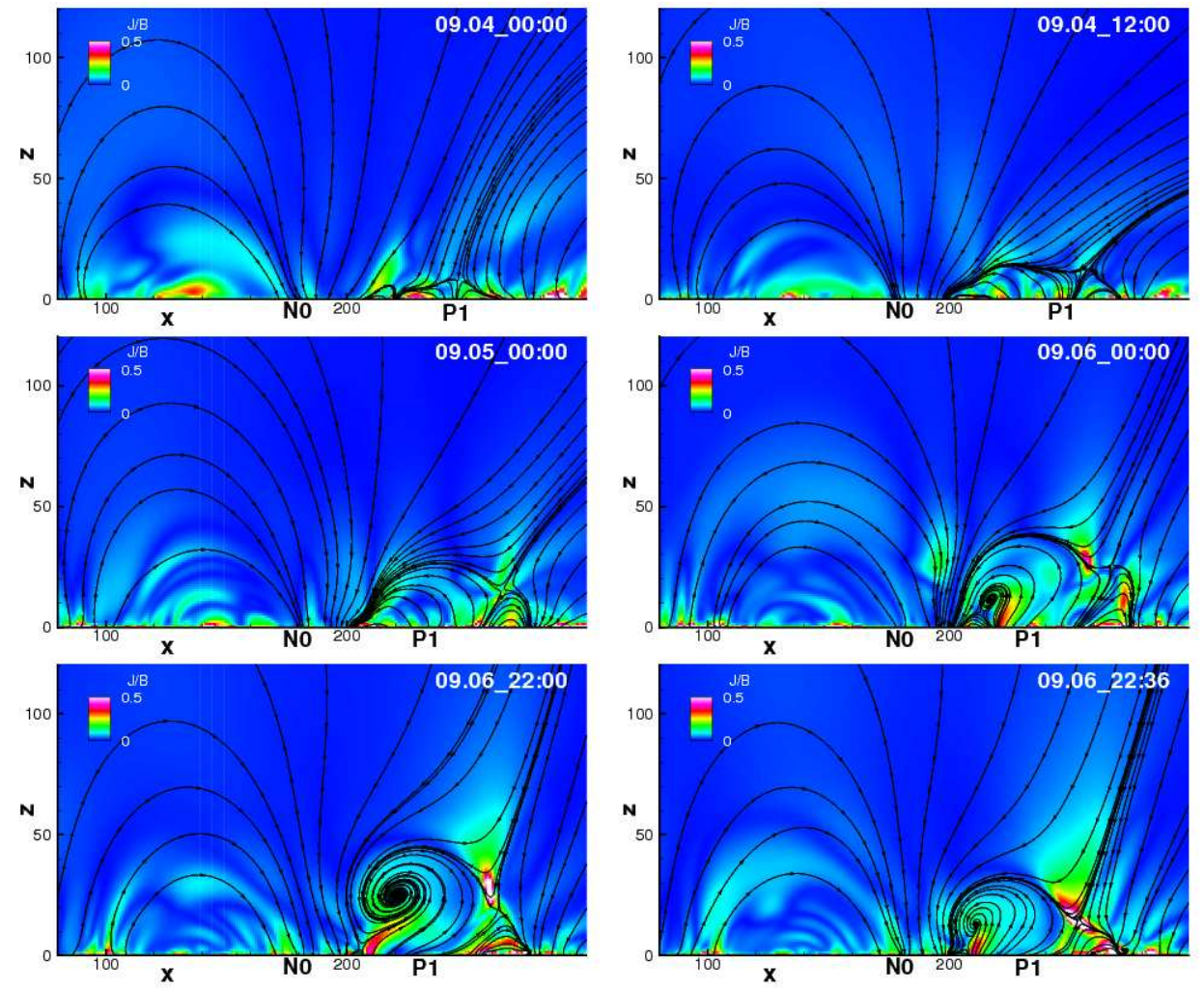}
  \caption{Evolution of a vertical slice data ($y=115$) from the 3D
    NLFFF volume. We choose such slice passing through the main polarities (
    P0, N0 and P1) for outlining the temporal evolution of the basic magnetic topology. The black
    lines with arrows are field-line tracing of the magnetic components $(B_x,B_z)$ on
    the $x$-$z$ plane, thus do not necessarily show the 3D field-line
    connectivity. Colored images of $J/B$ is plotted to show roughly
    the locations of magnetic separatries and QSLs.}
  \label{slice}
\end{figure*}

\subsection{Basic Magnetic Configuration}

\Fig~\ref{basic_topology} shows the basic configuration of the
pre-eruption field. It reveals that a sigmoidal field is embedded in a
typical spine-fan topology of a coronal null point above P1. The
magnetic null is located in the northwest (the direction of the eruption)
of the sigmoid at a height of 13~Mm. In the figure we plot several
representative field lines touching the null, which form a remarkable
X-point configuration at the null. Field lines going through the null
are of two types called spine and fan; the spine consists of two
singular field lines, the inner and outer spines, belonging to two
different connectivity domains which are divided by a dome-shaped
separatrix surface formed by the fan lines (compare \Fig~1 of
\citet{Pariat2009} who gave an idealized model of null
topology). Naturally, the fan surface intersects with the chromosphere
in a closed circle. The circular flare ribbon is an evidence of
the reconnection occurring at the null, and is produced by
reconnection-accelerated particles along the fan lines down to
the chromosphere. The outer spine extends to the northwest,
and it appears to be closed connecting out of the
extrapolation box, as shown by the large-scale field lines in
\Fig~\ref{fig:1.0}. The correspond remote
brightening at the outer spine footprint is observed in the northwest
of the AR (with a distance of roughly 200~arcsec) at about
22:23~UT, almost at the same time as the appearance of the
circular ribbon.
We note that there are also possibly some field lines around the
outer spine even connected to P0, since another remote ribbon
appears at P0 near the flare peak time \citep{Dai2013}.

In \Fig~\ref{slice}, we plot the evolution of a vertical cross section
of the field to show the formation of the basic topology. The null
topology on the two-dimensional (2D) slice displays a X-point configuration. As P1
emerges, the null forms and lifts up as the field lines expand, along
with successive expansion of the dome-like fan domain. In this domain
the field lines connect P1 to N0, and out of the domain the field lines
connect N0 to P0 or positive polarities out of the FoV, while N1
merges with N0. In this process, small reconnections at the null
account for the connectivity interchange between field
lines in and out of the closed fan domain. This reconnection usually
manifests itself by small flares with circular shape tracking the
chromospheric footpoints of the fan lines. This basic magnetic
topology forms at the end of the first day while the field is still
near potential
(also see Section~\ref{sec:energy} for the analysis of the
volume energy). After that a FR forms gradually above the main PIL of P1/N0, as
indicated by the spiral of the 2D-projection field lines, which are
a manifestation of the poloidal flux in the rope. Approaching the eruption,
the 2D spiral expands greatly along with an increase of the current,
as more and more flux is transferred into the rope. In \Fig~\ref{slice},
a distinct variation of $J/B$ (i.e.,
the currents normalized by magnetic field strength) can be seen around the spiral flux.
As shown by \citet{Savcheva2012a,Savcheva2012b} and \citep{Aulanier2010}
such current concentrations match the main QSL that separates the FR from its surroundings.


Although the magnetic field is reconstructed statically, the
successive results reproduce vividly the emerging process. It is worth
noting that \Fig~\ref{slice} shows field constrained by data and
not theoretical models of flux emergence in previous works \citep[e.g.][]{Shibata1998}.

\begin{figure*}[htbp]
  \centering
  \includegraphics[width=0.9\textwidth]{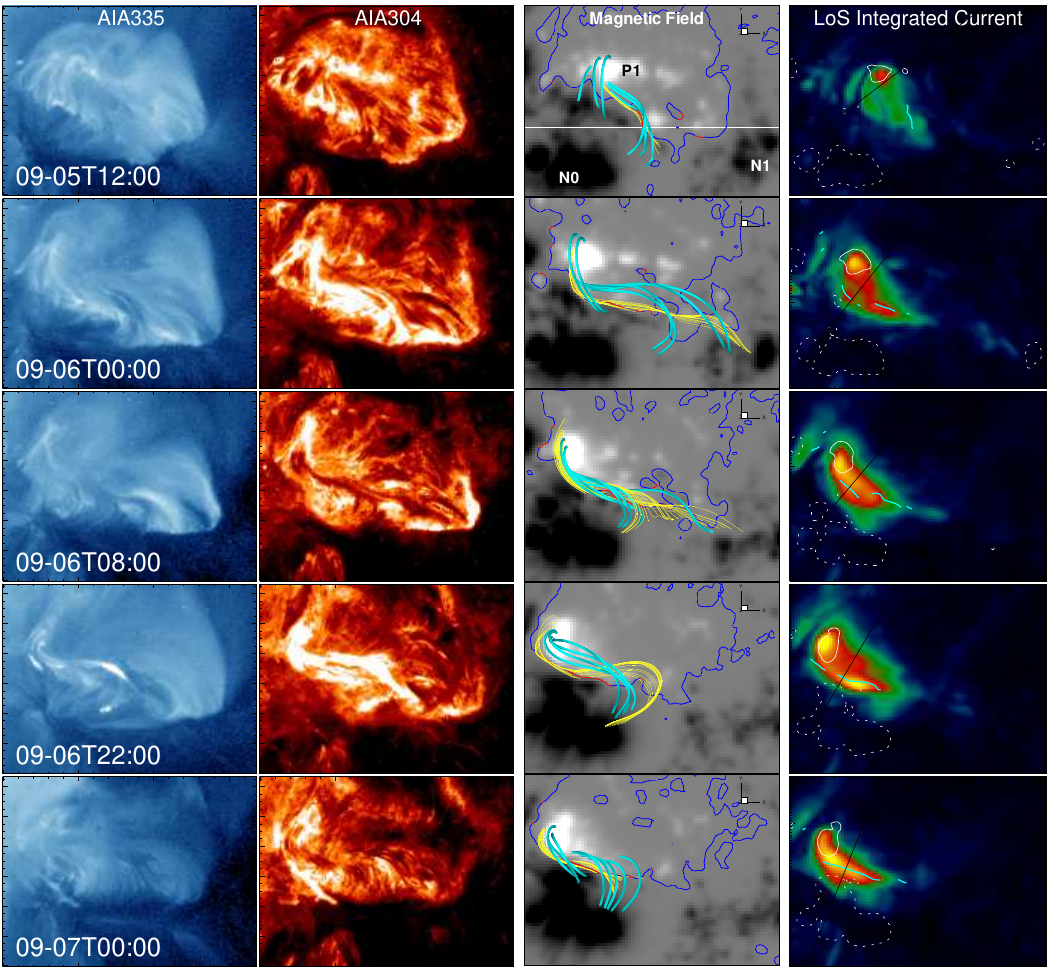}
  \caption{Evolution of the core field. Plots from left to right are
    respectively AIA-335 and AIA-304 images, sample coronal field
    lines, and LoS integrated current. In the magnetic field plots,
    the blue contour lines is the PIL and the red portions denote the
    BPs; field lines in yellow are traced through the BPs, forming the
    BPSS; the cyan lines are some sample field lines closely above the
    BPSS. In the plots of integrated currents, brighter color
    indicates more intense currents; the contour lines are for the
    $B_{z}$ of $\pm 1000$~G; the cyan segments are the BPs; and the
    black lines denote the locations where we slice the central cross
    sections of the FR.}
  \label{sigmoid_evolution}
\end{figure*}

\begin{figure*}[htbp]
  \centering
  \includegraphics[width=0.9\textwidth]{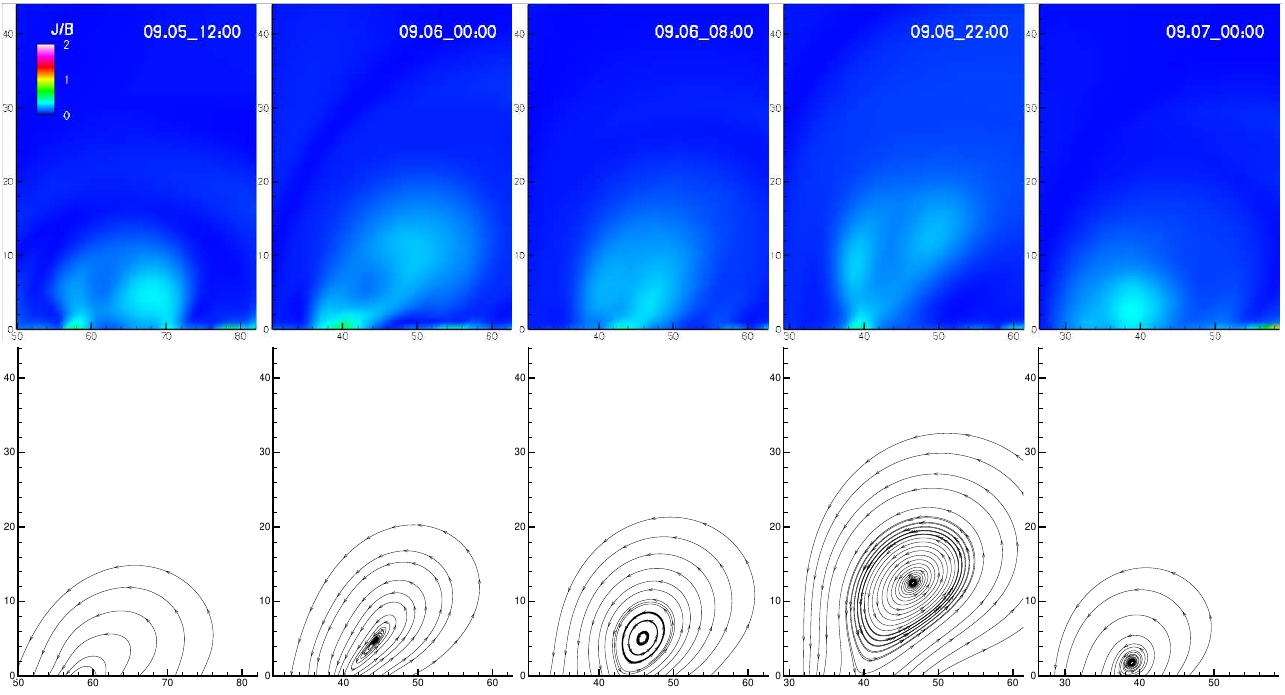}
  \caption{Evolution of the central cross section of the FR. Locations
    of these slices are denoted by the black lines in the last column
    of \Fig~\ref{sigmoid_evolution}. The top panels show currents
    normalized by the field strength, and the bottom panels show the
    field-line tracing of poloidal flux of the FR, which forms helical
    field lines centered at the axis of the FR.}
  \label{censlice}
\end{figure*}

\begin{figure*}[htbp]
  \centering
  \includegraphics[width=0.8\textwidth]{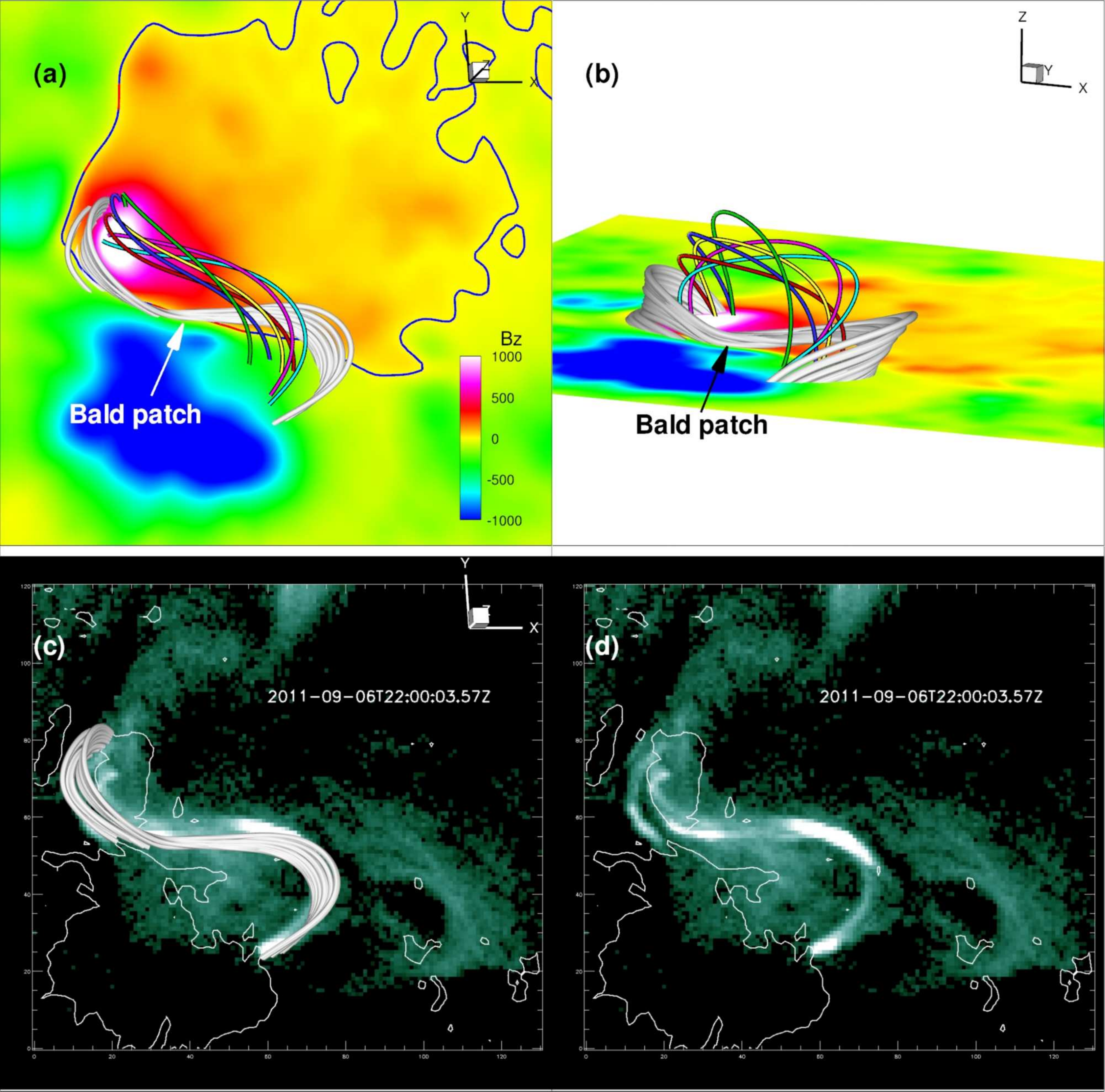}
  \caption{Observation and NLFFF reconstruction of the sigmoid field
    at 22:00~UT on September 6. The white thick lines are the BPSS
    field lines, which graze the photospheric surface at the BP. The
    inner body of the FR is shown by the color field lines. In
    particular the yellow line is the FR axis, which is traced through
    the center of the helical flux in the central cross section of the
    FR as shown in \Fig~\ref{censlice}. The field lines are shown in both
    the {\SDO} view (a) and a side view (b). Panels (c) and (d) are the AIA-94
    image of the sigmoid and the BPSS field lines are overlaid on the
    AIA observation in panel (c). Contours of $\pm 500$~G for $B_z$
    are also overlaid on the AIA images.}
  \label{sigmoid_NLFFF}
\end{figure*}

\subsection{The Core Field}
\label{sec:formation}

Using an idealized numerical simulation, \citet{Aulanier2010} showed that
in a sheared bipolar region, the building-up of a FR begins with the
appearance of BP. In this event, the BP appears first at about
12:00~UT, September 5. In \Fig~\ref{sigmoid_evolution}, we plot four
snapshots of the buildup process of the FR and a snapshot after its
eruption. The magnetic field lines are compared with the observations
in a high-temperature channel (AIA-335, 2.5~MK) and a low-temperature
channel (AIA-304, 0.05~MK). In the magnetic field plots, the yellow
lines are traced through the BP (red parts of the PIL), forming the
BPSS; the cyan ones are some sample field lines closely above the
BPSS. The integrations of the current density $J=|\crlB|$ along $z$
(approximately in the line-of-sight (LoS)) are plotted to show the main body of the
FR. We note that in the corona, besides the volume currents, currents
also develop in the form of thin sheets along the BPSS and other
separatrix and QSL, but these current sheets ought to be very
thin and beyond our grid resolution. Furthermore these current sheets
form only in the presence of footpoint motions, thus it is difficult
to recover them by a static extrapolation model
\citep{Savcheva2012a}. As a result, in our model, only the FR's volume
currents contribute to the LoS integrated currents. In
\Fig~\ref{censlice} we further plot a vertical slice of the NLFFF data
for each snapshot (locations of these cuts are overlaid on the plots
of the LoS integrated currents in \Fig~\ref{sigmoid_evolution}). These
slices are cut through the middle of the current concentrations in a
direction perpendicular to the central PIL, thus corresponding to the
central cross sections of the developing FR. In these cross sections,
as in \Fig~\ref{slice}, the poloidal flux of the FR forms a helical
shape and the helical center can be regarded as the apex of the FR
axis. In \Fig~\ref{censlice} we also show the distribution of $J/B$ which can help
us identify sites of possible magnetic separatries and QSLs \citep[e.g.,][]{Fan2007,Pariat2009}.
As shown in the plots, $J/B$ roughly enhances the boundary
of the rope and its ambient flux, i.e., the BPSS.

It is interesting to find that the EUV structures are matched very
well by the BPSS and nearby field lines rather than the LoS integrated
currents (i.e., the FR's main body). As can be seen, the hot
structures and the cold filaments are co-spatial with each other
and are both resembled well by the BPSS field lines. The reasons are
twofold: on the one hand, as aforementioned, BPSS are preferential sites where thin sheet
of current forms due to persistent photospheric motion. Reconnection occurs continuously at the BPSS and
produces the high-temperature emission. The volume currents in
the FR, although strong, are extended, thus its dissipation is much
slower than those BPSS currents, and produce no enhanced emission. As a result, the hot emission mainly corresponds to the BPSS and
related field lines. For the same reason, high-temperature EUV emissions are also
enhanced along the fan separatrix of the null point, making the separatrix rather
clear in its evolution (see \Fig~\ref{fig:AIA+NLFFF_evolve}). On the other
hand, the BPs are places where field lines are tangential to the
photosphere and thus concave upward. Thus, the field lines just above
the BPs form dips where the filament matter can be sustained, which
explains why the filaments seen in AIA-304 are closely co-spatial with
the BPs \citep[e.g.,][]{Aulanier1998}.

The BPSS field lines coincide indeed well with the sigmoid.
In \Fig~\ref{sigmoid_NLFFF}, as an example, we plot a set of BPSS field lines (the
thick white lines) and overlay them on the AIA-94 image of 22:00~UT,
September 6, when the sigmoid structure was observed most clearly just
prior to its eruption. The sigmoid has a thin and enhanced forward-S
shape (indicating a right-hand twist) in the AIA-94 wavelength
(6.3 MK), and is also well shaped but more diffuse in SXR taken by
Hinode/XRT. Perfect alignment of the BPSS field lines in the
S-shape of the sigmoid can be seen. We also plot field lines near the axis of the FR
(the thick colored lines) to compare its inner body with
the sigmoid. These field lines are weakly twisted and also exhibit an
S-shape slightly, but they clearly do not match the sigmoidal emission
in either location or shape.
Although here we cannot recover the BPSS current sheet, MHD simulation
indeed shows that current sheet spontaneously develops along the BPSS
once the coronal field is driven by photospheric motions
\citep{Pariat2009, Aulanier2010}. This supports the BPSS model for
this sigmoid in that the current sheet forming along the sigmoid-shaped BPSS
produces the enhanced emission of the EUV and SXR sigmoid
\citep{Titov1999, Gibson2004, Gibson2006, McKenzie2008}.

\begin{figure*}[htbp]
  \centering
  \includegraphics[width=0.8\textwidth]{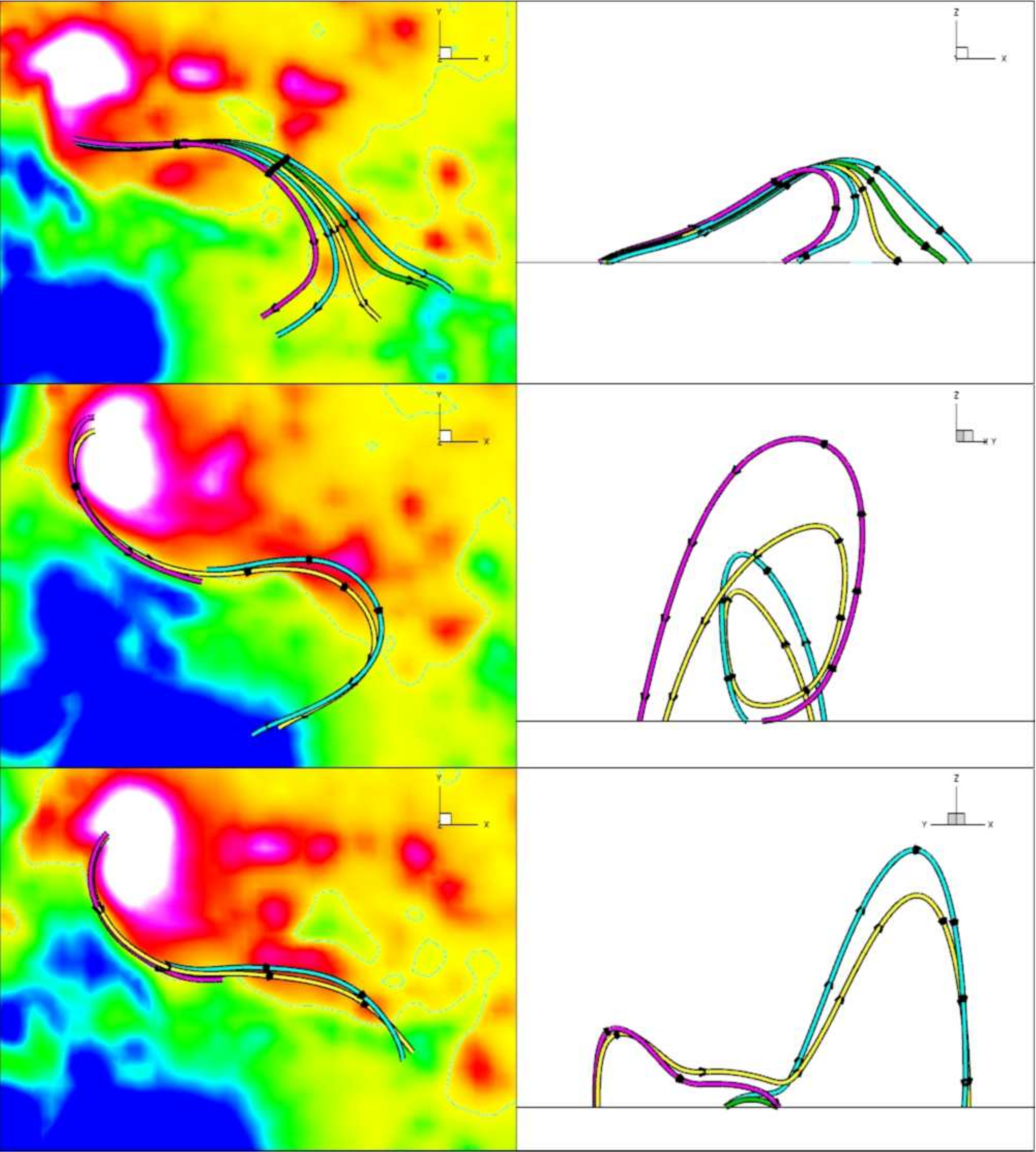}
  \caption{Illustration of different magnetic reconnections involved
    in the FR formation. For each panel, the field lines are selected from a
    NLFFF model at a given time, thus they do not represent the
    dynamic evolution of field lines during reconnections, but help us
    understand how field line connectivity is changed by the
    reconnection. Top: the slipping type of reconnection. The field
    lines are rooted extremely closely in their north footpoints but diverge
    largely in their south footpoints since they are in a QSL. The
    continuous set of field lines illustrates the slipping of a single
    field line along the QSL, with its south footpoint moving from
    right to left during the reconnection. Middle: flux-cancellation
    reconnection. Footpoints of two inverse J-shaped field lines (the
    pink and cyan lines) are brought so closely toward the PIL that they
    connect, forming a long S-shaped field line (yellow). Bottom:
    coronal tether-cutting reconnection. Two inverse J-shaped arcades
    (pink and cyan) are sheared past each other with their arms;
    reconnection in the corona occurs at the closest point between
    their arms, results in a long S-shaped FR field line (yellow) and
    a short sheared arcade below (green). In the side views shown in the right panels,
    the vertical axis is stretched for a better visibility of different
    field lines.}
  \label{reconnection_example}
\end{figure*}

\begin{figure}[htbp]
  \centering
  \includegraphics[width=0.45\textwidth]{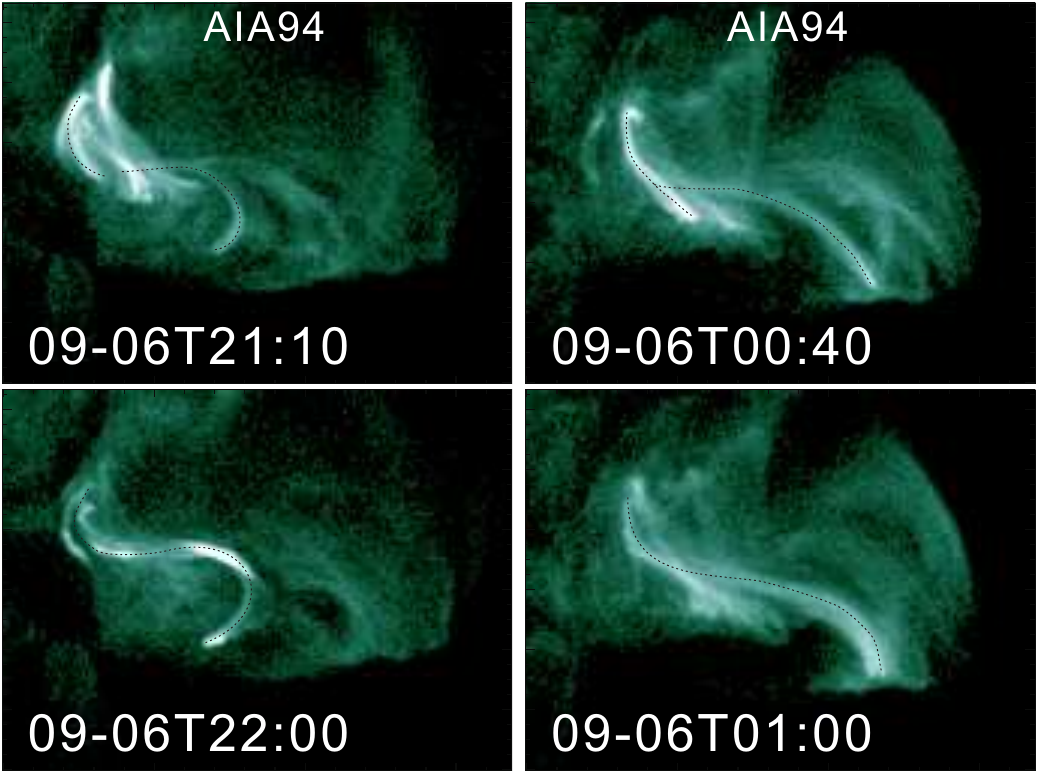}
  \caption{Left: Observation of two inverse J-shaped arcades which merge at
    their inner footpoints into a single S-shaped loop, as in the flux cancellation
    cartoon. Right: Observation of
    two inverse J-shaped arcades, which are sheared past each other
    with their arms, reconnect into a long S-shaped loop, as in
    the coronal tether-cutting reconnection. The dashed curved lines
    overlaid on the images denote the loops.}
  \label{obser_reconnection}
\end{figure}

\subsection{Formation of the Flux Rope}

Following the time evolution of the magnetic field, we see
a transition of the initial potential arcade to a double-J-shaped sheared
arcade and further to a S-shaped FR, along with the increasing of the
volume current. This process is driven by the diverging motion between
P1 and N1, the shearing motion between P1 and N0, and the rotation of
P1. Clearly the FR does not emerge bodily with magnetic flux from
below the photosphere, but is built up in the corona by a combination of
different kinds of magnetic reconnections, which is discussed
below. In \Fig~\ref{reconnection_example} we select field lines in the
static model to illustrate these reconnection processes. Although
these plots do not represent the dynamic evolution of field lines
during reconnections, they help us understand how the field line
connectivity is changed by the reconnections.

As shown in \Fig~\ref{sigmoid_evolution} from 00:00~UT to 22:00~UT on
September 6, some very long arcades (including some field lines in the
BPSS) connecting P1 and N1 develop J-shaped lines as their south
footpoints slip from N1 to N0. Such transformation can be attributed
to the slipping type of reconnection, in which the field line changes
its connection quickly along QSL \citep{Aulanier2006}. The presence of
a QSL here can be inferred by inspecting the field lines that are rooted
closely in P1 but diverge largely in their south footpoints, as shown
in the top panel of \Fig~\ref{reconnection_example}. A computation of
the squashing factor can further locate precisely the QSL
\citep{Savcheva2012a, Savcheva2012b}, which is omitted here.
This slipping reconnection is driven by the diverging motion of
the new-emerging N1 and P1, which stresses the field, and results in
a narrow current layer and reconnection in the QSL. For these field
lines, the slipping of their footpoints from N1, a polarity getting
farther away from P1, to the much nearer one, N0, is a natural result of
the magnetic relaxation to a low-energy state. Due to this slipping
reconnection, the final FR forms not between the new emerged bipole
P1/N1, but with its south leg rooted at the pre-existing polarity N0. Such a
reconfiguration further supports that the FR does not emerge bodily
but forms in situ in the corona.

Flux cancellation is a basic process that transforms the double-J-shaped
field lines to a S-shaped FR \citep{Ballegooijen1989}. Actually, the BP
forms as a result of the flux cancellation and the first appearance of
BP also marks the beginning of flux-cancellation reconnection. The
flux-cancellation reconnection is driven by the photospheric converging
motion at the PIL. When the footpoints
of two inverse J-shaped arcades are brought closer and closer to each other by convergence flows towards the PIL, reconnection between these
footpoints occurs on the photosphere (see the middle panels in
\Fig~\ref{reconnection_example}). It results in a long field line,
typically of the sum length of the two J-shaped arcades, and a much
shorter loop which submerges quickly, observed as an annihilation of
inverse flux elements toward the PIL. During the reconnection, the
long field line is instantly a BPSS field line that touches the
photosphere at the reconnection point (also the BP point), and quickly
detaches from the photosphere to be a field line inside the FR. Since
the reconnection point can be considered to be on the photospheric
surface, this process is also called ``photospheric tether-cutting''
reconnection \citep[e.g.,][]{Aulanier2010}, to distinguish it from the
standard tether-cutting reconnection that occurs in the corona
\citep{Moore2001}. An example of the observation in AIA-94 is shown in
the left panels of \Fig~\ref{obser_reconnection} where two J-shaped
arcades merge into a single S-shaped loop. As more and more sheared
arcades reconnect in this way, flux in the rope increases and the FR's
axis ascends higher and higher (see \Fig~\ref{censlice}). At
the same time, the length of the BP grows (see \Fig~\ref{fig:linevol},
(d)), and also the BPSS.
Although the BP breaks into fragments sometimes, it persists throughout
the whole formation phase and remains even through
the eruption. As reconnection occurs at the BP during its whole life time, the building
of the FR is conducted mainly through the flux cancellation.

It has been shown that in the case of sigmoidal FR forming in a decaying
bipolar AR, the BP bifurcates, shrinks and
eventually disappears, the BPSS transforms gradually to a QSL with a
HFT (i.e., a magnetic X-line-type structure) at the center,
and the standard coronal tether-cutting reconnection sets
in below the FR, elevating its main body off the photosphere \citep{Aulanier2010,Savcheva2012a,Savcheva2012b}.
In our event, the fragmentation of the BP line does not correspond to a
systematic bifurcation of the BPSS, since the BP segments do not shrink and disappear and
the FR main body is attached to the photosphere during its whole formation
phase. Even though, one cannot exclude the possibility of
the standard tether-cutting reconnection occurring above the central
part of PIL between the BP segments. The field lines near these
places, as illustrated in \Fig~\ref{reconnection_example}, resemble
the configuration of the coronal tether-cutting reconnection: two
inverse J-shaped arcades are sheared past each other with their arms, a
S-shaped FR field line lies just above them and a short sheared arcade
below. In the tether-cutting reconnection, the two J-shaped arcades
reconnect in the corona at the closest point between their arms above
the photospheric PIL, which converts J-shaped arcades into the
S-shaped field line and forms also the small arcade below the FR. In
\Fig~\ref{obser_reconnection}, we also identify an observation example
that agrees well with such transition of the arcades. We thus suggest that the standard tether-cutting reconnection also contributes to the building
up of the FR along with the flux cancellation.

During the buildup process, more and more flux is fed into the FR,
which expands upward slowly and stretches out the envelope field, as
clearly shown in \Fig~\ref{censlice}. The FR is pushed to the
northwest significantly because the magnetic flux (and thus the
magnetic pressure) of the south polarity N0 is much stronger than that
of P1. Up until its eruption, the FR reaches a height
of 14~arcsec (10~Mm) and deviates from the vertical by approximately
$30^{\circ}$, and most flux of the FR is on the right side of the
PIL as seen in the cross section. During its whole life time,
the FR is attached to the photosphere at the BP, while its cross
section almost develops into an inverse tear drop shape just prior to
the eruption.
Correspondingly, the vertical cut of the currents in the BPSS also
develops from a U-shape to a nearly V-shape.

\subsection{Evolution through the Eruption}
\label{sec:across}

A distinct release of the non-potentiality can be seen through the
eruption. Both \Figs~\ref{slice} and \ref{censlice} show that through
the eruption the helical core contracts significantly and its envelope
flux relaxes downward. As a result, the core field lines become much shorter
and thus more potential, as shown in \Fig~8, and also the volume current decreases and
becomes more compact. Such reformation is consistent with the enhancement of the transverse field
on the photosphere.
Although the BP shows no substantial decrease,
the BPSS shrinks significantly toward the PIL, as its long S-shaped
portion disappears and a much thinner shape remains. The post-flare arcades now straddle orderly over
the reformed BPSS, which also indicates that a FR lies below the
post-flare loops. The continued presence of FR attached to the
photosphere during and after eruption might be attributed to a partial
expulsion of the pre-eruption FR \citep{Gilbert2001,
  GibsonFan2006}. In such cases the flare reconnection occurs within
the FR and breaks the FR into two parts, of which the upper one
escapes while the lower remains. We discuss this point further along
with the mechanism of the eruption.

\begin{figure*}[htbp]
  \centering
  \includegraphics[width=0.8\textwidth]{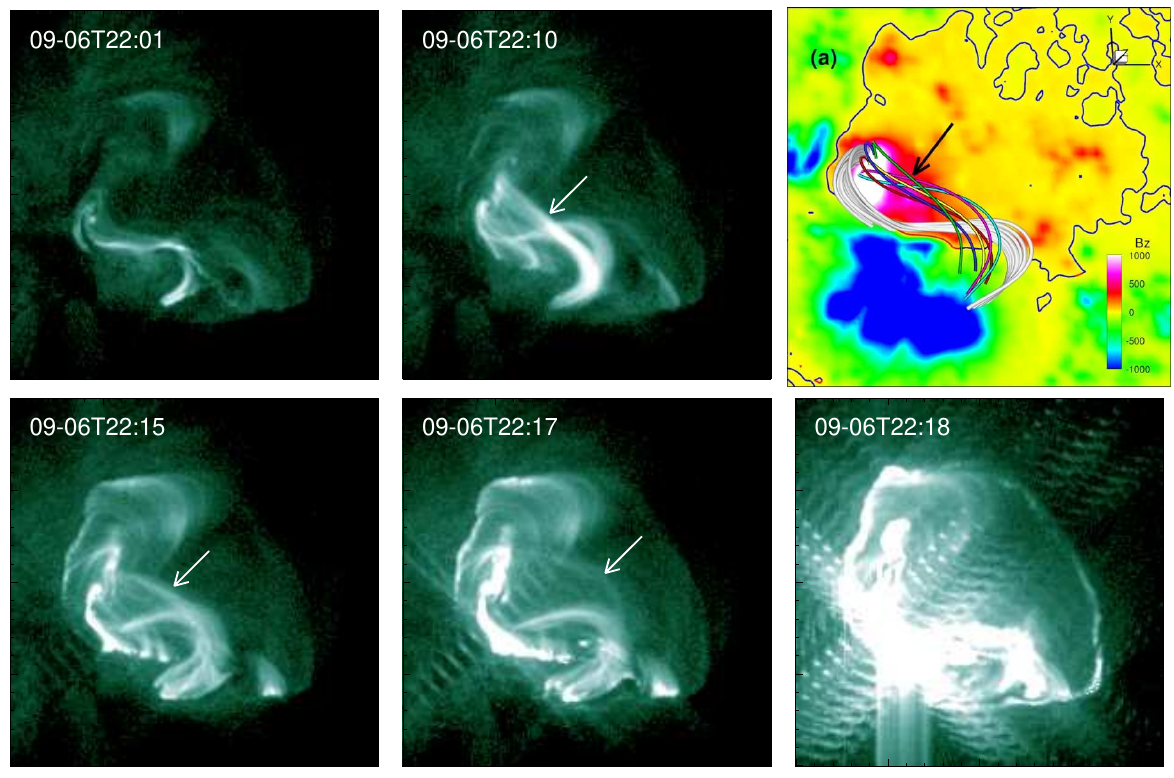}
  \caption{Observation in AIA-94 of the initiation process of the
    eruption. The arrows in the AIA images denote a group of erupting loops which
    are brightened progressively until 22:12~UT and then expand rapidly toward the northwest
    and become invisible after 22:17~UT. After that a
    remarkable circular flare ribbon appears.
    The panel (a) of \Fig~\ref{sigmoid_NLFFF} is put here to compare with
    the observation at 22:10~UT. Although the sigmoid evolves from 22:00~UT to 22:10~UT and
    thus deviates slightly from the BPSS field lines at 22:00~UT, the resemblance
    between the erupting loops and the field lines closely around the FR axis clearly suggests
    that the erupting loops correspond to those near-axis field lines.
    {\it An animation of this figure is available}}
  \label{early_eruption}
\end{figure*}

\begin{figure*}[htbp]
  \centering
  \includegraphics[width=0.9\textwidth]{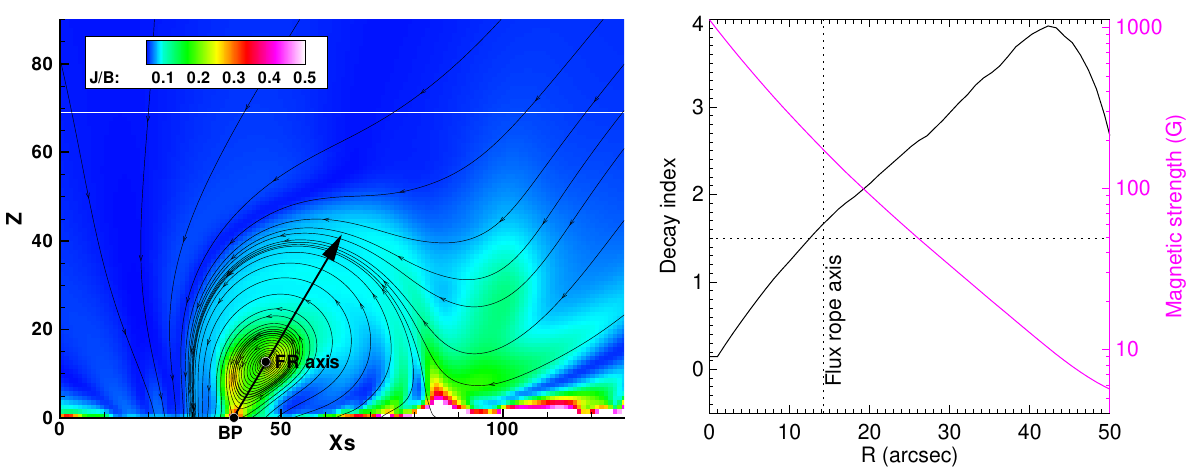}
  \caption{Left: central cross section of the core field at 22:00~UT
    on September 6. The background shows the current normalized by the
    field strength. The two black dots denote the locations of BP and
    FR axis, respectively. The decay index is computed along the
    arrowed line pointing from the BP to the FR axis. Right:
    distribution of decay index and strength of the restraining field
    with the distance $R$ starting from the BP. The horizontal dashed line denotes
    decay index of 1.5 (the critical threshold for TI) and the
    vertical dashed line denotes the location of the FR axis.}
  \label{decay_2200}
\end{figure*}

\begin{figure}[htbp]
  \centering
  \includegraphics[width=0.45\textwidth]{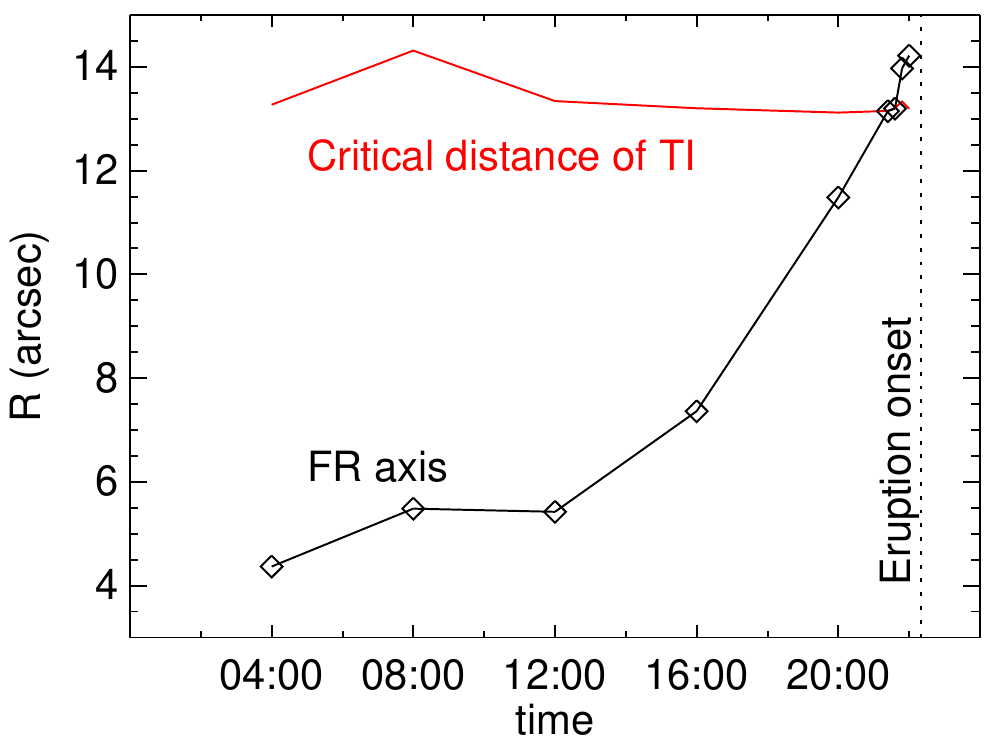}
  \caption{Time evolution for the distance from the BP to the FR axis
    and the critical distance from the BP to the TI domain during the whole day of September
    6. The vertical dashed line denotes the time of eruption onset.}
  \label{decay_evol}
\end{figure}


\section{Initiation Mechanism of the Eruption}
\label{sec:eruption}

In this section we investigate what mechanisms are involved and what roles
they play in causing the eruption based on observations and
coronal fields at the onset of the eruption. The magnetic field evolution
of the eruption process will be analyzed in detail in the second paper.

\subsection{Erupting Loop}

\Fig~\ref{early_eruption} shows the AIA-94 observation of the early
phase of the eruption. A significant brightening of the sigmoid is
triggered at 22:00~UT and the sigmoid evolves with a clear expansion of
its west hook. Immediately afterwards, a group of loops overlying the sigmoid
(marked by arrows in the AIA-335 images) is also
lit up gradually, being even brighter than the sigmoid, and expand very slightly until 22:12~UT. After
that these loops accelerate rapidly toward the northwest,
while they progressively faint until being invisible, and are followed by the
appearance of a circular flare ribbon at about 22:17~UT.
Such erupting loops have also been observed in some other sigmoid eruptions
\citep[e.g., the linear bar-like features reported by][]{McKenzie2008, Green2011},
but their relation with the erupting FR is not very clear. They have been usually speculated as the erupting FR
itself \citep[e.g.][]{Moore2001,McKenzie2008,Liu2010,Green2011}. \citet{Aulanier2010} suggest
another interpretation: they are not associated with the erupting FR
but with a current shell that develops within expanding field lines above the rope.
In the present case, the observation of such erupting loop is much more defined than
in those previous studies. By a direct comparison with the coronal field at 22:00~UT,
we can clearly identify such erupting loop as indeed part of the FR, and in particular, it
corresponds to those field lines near the rope axis (see the close resemblance between the
erupting loop observed at 22:10~UT and the near-axis field lines shown in top-right panel in \Fig~\ref{early_eruption}). The accelerated rising of this erupting loop thus indicates the
initial eruption of the FR itself, and it sheds
important light on our following study of the initiation mechanism.

\subsection{Magnetic Breakout vs. TI}

In this event, we can first exclude the KI since the twist of the FR,
less than one winding of field lines, is too weak to trigger KI. Also
we have not observed a clear rotation of the erupting loop (i.e., the FR axis)
as it rises, which would otherwise occur in the KI of a FR. The
sheared core embedded in a null point topology resembles the
lateral breakout model except that the FR has already formed before
eruption.
The growing magnetic pressure of the FR could cause its closely
overlying flux to expand outward and stress the null point related
separatrix, which eventually triggers the breakout reconnection.
Furthermore, the quasi-circular flare ribbon supports such a null-point reconnection.
However, observations of the erupting loop (\Fig~\ref{early_eruption}) shows
that the accelerated expansion of the FR starts before the appearance of the circular
ribbon,
which suggests that
the eruption cannot be triggered by the breakout reconnection.

The TI might apply to the FR-overlying field system
of the core field because such flux system was formed long before the
breakout eruption. We note that the FR in this event does not fully
develop a nearly semi-circular configuration with its main body
elevated off the photosphere (except for the two legs anchored at the
bottom), as in the case of a standard TI for coronal FR. Nevertheless,
we can study whether the central flux of the rope reaches
the domain of TI by calculating the decay index of the restraining
field near the apex of the FR axis.

In the central cross section of the FR, we compute the decay index
along the direction pointing from the BP to the FR axis, as
illustrated in \Fig~\ref{decay_2200}. Since the FR stretches
in this direction in the build-up phase, we show that it expands in the same
direction when the TI occurs. Note that the TI only applies to the
closed flux domain under the fan separatrix of the null, thus the decay
index is calculated within this domain. The restraining field, also
referred to as the external field, can be approximated by the
potential field with the same vertical magnetogram of the NLFFF
\citep{Fan2007, Aulanier2010}. Furthermore, because the field parallel
to the rising direction actually does not contribute to the inward
confining force, the decay index is computed for only the
perpendicular component of the potential field
\citep{Cheng2011a,Nindos2012}. We then calculate the decay index
\begin{equation}
  \label{eq:decay_index}
  n(R) = -\frac{R}{B}\frac{\partial B}{\partial R}
\end{equation}
where $R$ is the distance from the starting point, i.e., the BP at the
bottom, and $B$ the perpendicular component. As plotted in the right
panel of \Fig~\ref{decay_2200}, the decay index climbs to a critical
value of 1.5 for TI \citep{Torok2007} at a distance of about
13~arcsec, and stays above 1.5 for the rest of the domain. Note that
the decay index does not increase monotonically but inflects near the
separatrix between the closed and opened fluxes,
possibly due to different variation profiles of the different flux systems.
The FR axis is located at a distance of about 14~arcsec, showing that it already runs into
the TI domain at 22:00~UT on September 6, a few minutes prior to the
eruption onset.

In \Fig~\ref{decay_evol}, we plot the time evolution of $R_{\rm
  axis}$, i.e., the distance of the FR axis from the BP, and $R_{\rm
  TI}$, i.e., the critical distance for TI, during the whole day of
September 6 before the eruption. With the build up of the FR,
the apex of the rope axis gradually approaches
the TI domain. This plot clearly shows that the
eruption occurs immediately (within a few minutes) once the FR axis
reaches the TI domain, which strongly suggests that the TI is the trigger
mechanism of the eruption. Also, as inferred from the time profile of $R_{\rm axis}$,
the FR rises very slowly with a speed of a
few tenths of a kilometer per second.
It demonstrates the quasi-static nature of the slow buildup process of the FR, which can indeed be modeled by the static extrapolation method.

On the other hand, the critical distance
of TI shows no significant variation because the potential field
changes very litte during the day. This is consistent with
the finding of \citet{Nindos2012} that the initiation of eruptions
does not depend critically on the temporal evolution of the decay index
of the background field. Indeed the photospheric flux
is modified very slightly by flux cancellation, being on the order of
$5\%$ (see \Fig~\ref{fig:linevol} (a)) during the day, and also the flux
distribution shows no significant variation (see \Figs~\ref{local_magram} and
\ref{sigmoid_evolution}). Therefore we can
further clarify that it is not the modification of the external
restraining field, by lowering either its critical distance for TI
or its magnetic tension force, that leads to the eruption.

\citet{Jiang2013apjl} demonstrated
the capability of reproducing the realistic solar eruptions using their data-constrained
MHD model, which achieved a perfect resemblance of the simulation of the filament eruption
with observations \citep[compare Figure 1 and Figure 5 of][]{Jiang2013apjl}.
But in that paper it is speculated that
the null-point reconnection triggers the TI. This is because the decay index there was
computed for the total magnetic field, which showed that the rope axis almost
but not yet reaches the domain of TI. Also the observation of the erupting loop
rising before the appearance of the circular ribbon was not noticed. We thus note that cautions are
needed in judging the TI using the decay index, because in the realistic magnetic configuration, both
the location of the FR axis and the strength of the external field are difficult to be precisely determined.

\subsection{Partial Expulsion of FR}

It is still worth mentioning the short time lag of few minutes
(from about 22:00~UT to 22:12~UT) between the rapidly accelerated
expansion of the FR and its entering into the TI domain. In this
time interval, observations (see \Fig~\ref{early_eruption}) show significant brightening of the sigmoid and loops
near the FR axis, but these FR loops expand very slowly, which is different from the standard TI case where the FR expands exponentially once it runs into the instability domain. The reason is
possibly due to the restraining effect by the photosphere at the BP,
since the instability sets in before a full detachment of the FR from the photosphere. Clearly, the flux-cancellation reconnection, which is
driven by the photospheric converging motion, is a process too slow to
account for the detachment of the FR in such a short time scale of minutes.
As shown in the \citet{Aulanier2010} simulation, the systematic
bifurcation of BP to HFT, i.e., the process of the FR detaching from
the photosphere, needs hundreds of coronal Alfv{\'e}n times. As a consequence,
in the short time scale of eruption, the BP actually plays a role of
line-tied restraint for the FR. Thus reconnection in the corona is
expected to occur in order for the FR to detach from the photosphere. It is very
likely that this reconnection occurs within the FR and results in a
splitting of the FR \citep{GibsonFan2006}. We speculate that by the combined effects of line-tying at the
BP and TI-driven expansion of the upper part of the FR, the FR is torn
into two portions with an X-line type reconnection formed in
between. The nearly tear drop shape in the cross section of the
pre-eruption FR is a sign of this tearing effect. After that, the
upper FR can freely expand as driven by TI. This reconnection
dynamically perturbs the BPSS and results in the enhanced heating
of the sigmoid and the rope. The reconnection further reforms the BPSS, and also
leads to a standard flare ribbon below the upper portion of the FR.
Almost at the same time, the circular ribbon
appears after the moderate expansion of the FR axis. It suggests that
the TI-driven expansion of the FR pushes its overlying flux quickly
and triggers the breakout reconnection at the null, which produces the
circular ribbon.

We summarize the scenario of the eruption as follows. The FR
is built up slowly into the TI domain while the FR body still touches
the photosphere; a combination of the TI-driven expansion of the FR
and the line-tying at the BP results in magnetic reconnection within the FR and
thus partial expulsion of the rope; meanwhile the TI-driven expansion of
the envelope flux of the rope triggers breakout reconnection at the null,
which further facilitates the eruption. The TI is the trigger and
initial driver of the eruption, and the magnetic breakout plays the role of
a further driver.


\begin{figure}[htbp]
  \centering
  \includegraphics[width=0.45\textwidth]{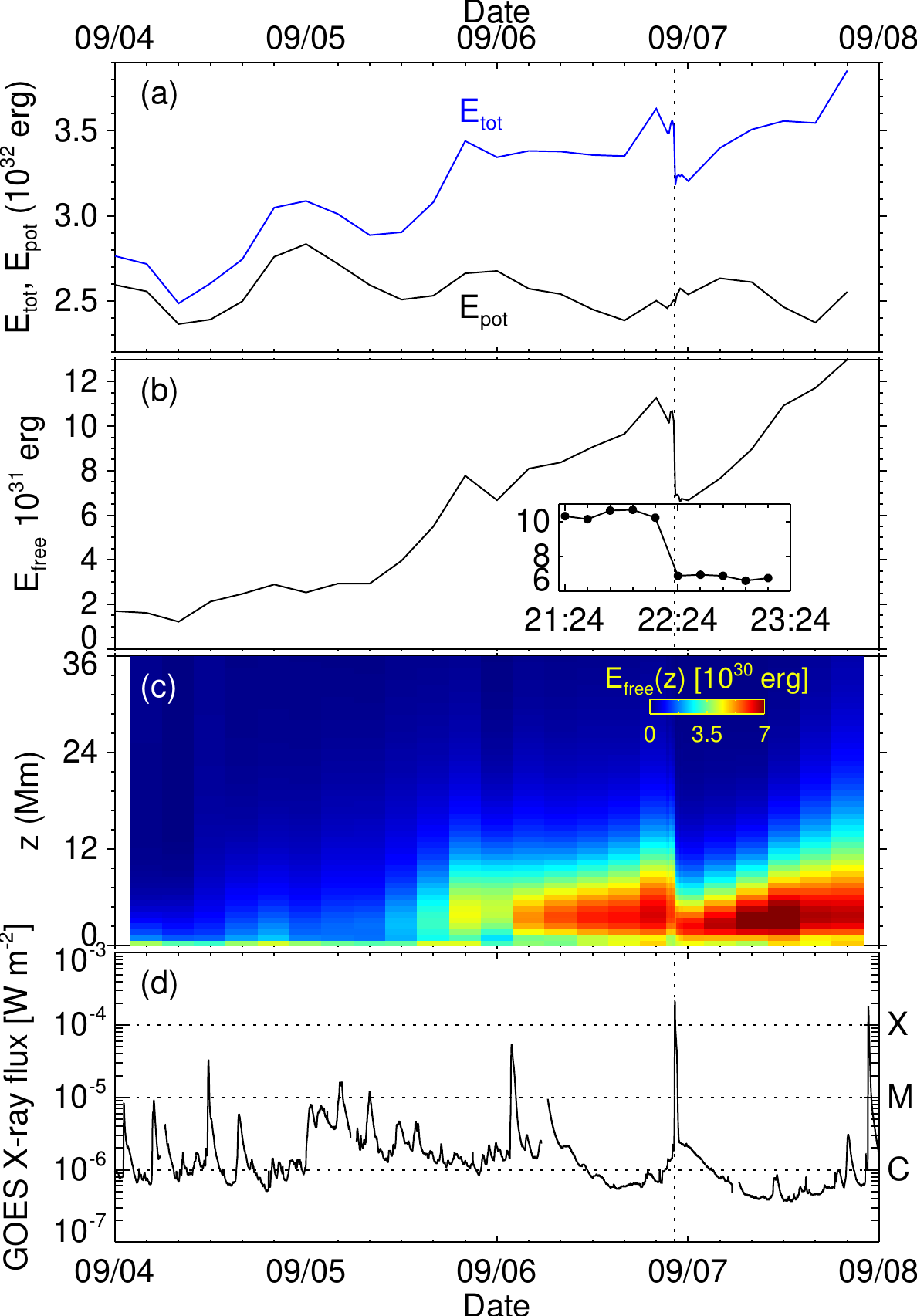}
  \caption{Evolution of magnetic energy: (a) Total magnetic energy $E_{\rm tot}$ and
    potential field energy $E_{\rm pot}$. (b) Magnetic free energy
    $E_{\rm free}$. (c) Vertical distribution of magnetic free energy
    $E_{\rm free}(z)$. (d) {\it GOES} soft-X ray flux.
    Inserts of (b) show result with 2~h around the flare peak time.}
  \label{erg_evol}
\end{figure}

\begin{figure*}[htbp]
  \centering
  \includegraphics[width=0.8\textwidth]{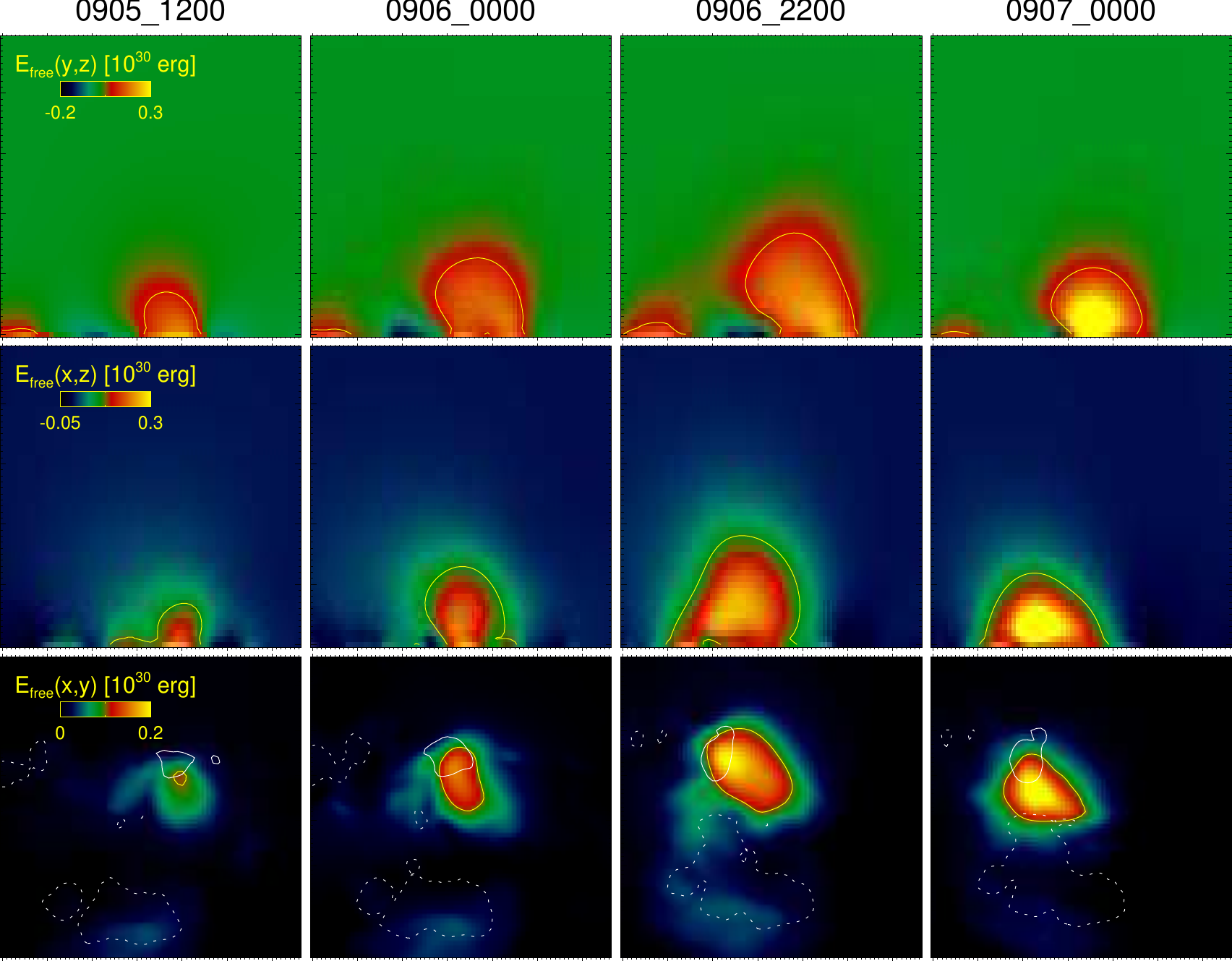}
  \caption{Evolution of the free energy distribution. From top to
    bottom are plots of free energy density integrated along the $x$,
    $y$ and $z$ axis, respectively. Three snapshots before the
    eruption and one snapshot after eruption are shown. The $z$ range
    for the top and middle panels is $z\in [0,50]$~arcsec, while the
    $x$ and $y$ ranges shown in these panels are the same as those of
    the bottom panels. The yellow contours are for value of $0.1\times
    10^{30}$~erg, and the white contours in the bottom panels are for
    $B_{z}$ of $\pm 1000$~G. }
  \label{erg_fre_dist}
\end{figure*}

\section{Magnetic Energy}
\label{sec:energy}

With the 3D coronal field, the volume energy can be computed. The
energy that can be released by eruptions is the free energy $E_{\rm
  free}$, i.e., the total energy $E_{\rm tot}$ subtracting the
potential energy $E_{\rm pot}$
\begin{equation}
  E_{\rm tot} = \int_{V} \frac{B^{2}}{8\pi} dV,\ \
  E_{\rm pot} = \int_{V} \frac{B_{\rm pot}^{2}}{8\pi} dV,\ \
  E_{\rm free} = E_{\rm tot}-E_{\rm pot}
\end{equation}
where $B_{\rm pot}$ is the potential field strength, and $V$
denotes the core field volume which is outlined by the boxes in
\Fig~\ref{magram_evolve} with a height of 50~arcsec. By calculations restricted
within this core region, we do not consider field energy irrelevant to
the eruption. Moreover most of the
free energy is contained in this core region \citep{Jiang2013NLFFF}.
The evolutions of these energies are plotted in
panels (a) and (b) of \Fig~\ref{erg_evol}. As expected, the potential energy evolves in
accordance with the photospheric magnetic flux, i.e., it increases mainly on
the first day and then decreases episodically due to flux cancellation.
The free energy increases only slightly by less than $1\times 10^{31}$~erg during the first day,
but climbs rapidly on the second day of more than
$\sim 5\times 10^{31}$~erg. This is consistent with the evolution
trend of the photospheric non-potentiality parameters, confirming that
most of the free energy is built up by line-tied
surface flows on the photosphere. The non-potential energy released on early September 6
is caused by an M5.3 flare around 01:30~UT. After that the free energy
increases moderately due to a small decrease of the potential energy
resulting from photospheric flux cancellation, while the total energy
remains almost the same. During the major flare at 22:20~UT, a step-wise drop of the total/free energy is captured by the full 12-min cadence of the NLFFF data. After this flare, non-potential energy is built up rapidly again due to persistent shear motions and is ready for the next eruption
near the end of September 7. The abrupt loss of energy during the major flare is intrinsically related with the abrupt change of the photospheric field. The amount of free energy
drop is $\sim 0.4\times 10^{32}$~erg, accounting for about 40\% of the
pre-flare free energy, which seems to be much lower than an estimated
value of $\sim 1\times 10^{32}$ by \citet{FengL2013} based on the sum of flare emission and
CME kinetic and potential energy for the studied event. This may due to many
reasons \citep{FengL2013,Sun2012}. One speculation is that
the NLFFF extrapolation overestimates the free energy of the post-flare field,
as it might be extremely dynamic and deviate from the force-free condition.
In addition, the smoothing of the original magnetograms may reduce the coronal energy content
by ignoring the small-scale flux elements and currents contained in the raw data.
There is also a possibility that the photospheric field measurements cannot resolve even smaller field structures that carry strong currents and a non-negligible amount of free energy.

We further study the distribution of the free energy by computing the
integrations of the free energy density along three axis lines, which
are respectively
\begin{eqnarray}
  E_{\rm free}(y,z) = dy~dz\int\frac{B^{2}-B_{\rm pot}^{2}}{8\pi}~dx,
  \nonumber\\
  E_{\rm free}(x,z) = dx~dz\int\frac{B^{2}-B_{\rm pot}^{2}}{8\pi}~dy,
  \nonumber\\
  E_{\rm free}(x,y) = dx~dy\int\frac{B^{2}-B_{\rm pot}^{2}}{8\pi}~dz.
\end{eqnarray}
\Fig~\ref{erg_fre_dist} shows their distributions for four
snapshots. By comparing the distributions of $E_{\rm free}(x,y)$ and
those of the integrated current in \Fig~\ref{sigmoid_evolution}, it
can be clearly seen that the free energy is stored largely
co-spatially with the current. This can be easily understood because
coronal free energy is actually stored in the current-carrying field
where the non-potentiality is strong. On the other hand, the free
energy generally does not concentrate within the sunspot umbras where
the magnetic flux is strongest. We note that in the plots there are
some small regions with negative values of integrated free
energy. This is physically valid since there is no restriction that
the energy density (and thus any sub-volume energy) must always be
greater than that of the potential field, although a non-potential
field must have a global energy content greater than that of the potential
field with the same surface flux \citep[e.g.,][]{Mackay2011}.

The evolution of the free energy distribution is consistent with the
evolution of the FR. The slow expansion of the FR is accompanied by the
expansion of the volume with free energy and the increasing of the
energy density. After the FR is partially expelled, the volume of free
energy shrinks and contracts downward abruptly and significantly. As a
result, the core of its distribution becomes so compact that the
density exceeds even that of the pre-eruption field, in spite of that
the total free energy drops during the eruption. In \Fig~\ref{erg_evol} (c),
we also plot the distribution of free energy along height $z$,
\begin{equation}
  E_{\rm free}(z) = dz\int\frac{B^{2}-B_{\rm pot}^{2}}{8\pi}~dx~dy
\end{equation}
for over four days. The increasing of the height of the free-energy domain toward
the major eruption can be seen. When the free energy becomes
significant, it is situated mostly near the height range of
$2-8$~Mm. Abrupt change during the X2.1 flare also appears
clearly, demonstrating a distinct downward contraction of free energy
distribution from above 4~Mm to a much lower height.

\section{Summary}
\label{sec:concld}

With the abundant data provided by {\SDO} and modern advanced numerical
models, we have unprecedented opportunities to examine in a
realistic and quantitative way many proposed mechanisms for solar eruptions, e.g.,
how the eruptive core field is built up, how the favorable magnetic
topology is formed, and when the system runs into a unstable regime
and erupts. In this paper
we studied a sigmoid eruption event in AR~11283 from its building up
to disruption for over three days, which involves a number of magnetic
processes and thus is attractive for our study. Based on a recently developed NLFFF
model \citep{Jiang2012apj,Jiang2013NLFFF}, we compute a time sequence of static
coronal fields to follow the slow buildup of the sigmoidal FR.
As opposed to most other NLFFF methods constrained by vector
magnetograms, our CESE--MHD--NLFFF code
can reproduce the structure of the evolving FR very well, as is demonstrated by the perfect
matching with the observations.
A detailed analysis of the fields compared with the \SDO/AIA
observations suggests the following scenario for the evolution of this
region.

Within the first day, a new bipole emerges into the negative polarity
of a pre-existing bipolar mature AR, forming a fan-spine topology of a
coronal null point on the separatrix surface between the two flux
systems. In the following two days, a FR is built up slowly in the
embedded core region through tether-cutting reconnections in both
the photosphere (i.e., flux cancellation) and corona, which is driven by photospheric shearing,
converging and rotating flows. In this process, BPSS forms between
the FR and its ambient field, and develops into a fully S-shape.
With more and more flux fed into the FR, the FR expands and the apex of the FR axis runs slowly into the TI
domain near the end of the third day. However the FR does not erupt
instantly since it is still attached at the bottom to the
photosphere. By the combined effects of the TI-driven expansion of the
FR and the line-tying at the BP, the FR is broken into two parts by reconnection within the rope. This reconnection dynamically perturbs
the BPSS and results in the transient enhanced brightening of the sigmoid. Then the upper portion of the
FR freely expands as it is driven by the TI, while the lower portion
remains. The fast expansion of the rope pushes strongly outward its envelope flux near the null point and triggers breakout reconnection at the null,
which further facilitates the eruption.

As to how a sigmoidal FR forms and erupts is still a subject of intense
debate, we summarize here the important results which are concluded from
the studied event but might also apply to other
events with similar magnetic configurations:

\begin{enumerate}

\item Magnetic flux emergence into an inverse-polarity preexisting
  field can form a fan-spine topology configuration with a coronal
  null on the separatrix surface between the two flux systems
  \citep{Moreno2008, Torok2009}. Flare ribbons with closed circular
  shapes trace the footpoints of the fan separatrix, and thus can usually
  be regarded as a signature of the presence of a null-related
  topology, along with the flux distribution of positive (negative)
  polarity surrounded by negative (positive) polarity.

\item A FR does not emerge bodily from below
  the photosphere, but forms gradually in the corona after the
  apparent new flux injection observed at the photosphere. The
  building up of a FR is largely driven by shear/rotation flows on
  the photosphere, which is possibly associated with the
  emergence. This is consistent with the numerical investigations of
  the flux emergence \citep{Magara2006, Fan2009} that a twist flux
  tube in the convection zone can not emerge bodily into the corona,
  but transport its twist by torsional Alfv\'en wave which manifests
  as the photospheric flows \citep{Longcope2000}.

\item Both flux cancellation \citep{Ballegooijen1989} and
  tether-cutting \citep{Moore2001} reconnections contribute to the in
  situ formation of FR in the corona from sheared arcades, but do
  not trigger the eruption \citep{Aulanier2010}. Such
  quasi-static evolution of the FR can be characterized well by a time
  sequence of static NLFFF models based on continuously observed
  magnetograms. The result also supports store-and-release CME models with FR
  existing in the corona prior to eruption, but not a by-product of
  eruption.

\item Comparison of the magnetic fields with AIA observations suggests that the
  prominent high-temperature EUV emission is largely produced by the
  current sheets developing along separatrix surfaces (and QSLs) but
  not by the extended volume current of the FR, because the dissipation rate of the
  extended currents is much smaller than that of the current sheets.
  In particular, the sigmoid is produced by the BPSS current sheet \citep{Titov1999}.

\item Although it has been shown that the TI triggers the eruption,
  the present case is different from that of the standard TI in which the
  FR is fully developed, i.e., is elevated off the photosphere away
  from a BPSS configuration for hours before eruption
  \citep[e.g.,][]{Aulanier2010,Savcheva2012b}. Here we demonstrate a case
  where the instability sets in before the FR is detached from the
  photosphere, and the photosphere can exert an additional restraining
  force to the FR at the BP. As a result, the FR does not erupt
  instantly even though its axis runs into the TI domain, and a splitting of the
  FR body is expected for the FR to expel partially
  \citep{GibsonFan2006}. Unlike the finding of \citet{Fan2007} that the partial
  expulsion of a FR (i.e., there are BPSS and FR remaining below post-flare loops) only occurs in the case of KI, we suggest that it
  can also occur in the case of TI.

\item An eruption is usually jointly produced by multiple mechanisms
  \citep[e.g.,][]{Williams2005}. For the studied event, in addition to the TI,
  the reconnection that splits the FR and the breakout reconnection that occurs at the null
  contribute to the final expulsion of the FR.

\item Magnetic fields experience abrupt changes through the eruption:
  the transverse field along the main PIL on the photosphere is
  enhanced; the long S-shaped BPSS shrinks significantly and reforms
  below the post-flare arcades, which is consistent with the enhanced
  photospheric field; the free energy is released mostly at a height
  of several Mms above the photosphere with a distinct downward
  compaction of its distribution. These results support the ``magnetic
  implosion'' conjecture. As a consequence, the non-potentiality of
  the photospheric fields might even increase after flare, and it is
  necessary to look at the 3D coronal field to disentangle these effects.

\end{enumerate}

Although the basic scenario of the AR evolution has been drawn,
questions remain for the dynamical evolution of the coronal field
during the eruption, in particular, how does the reconnection occur within
the FR, how does the BPSS evolve during this reconnection, and how is the
breakout reconnection triggered. A solution to these requires an MHD
simulation which is beyond the scope of this paper, and will be
investigated in detail using the recipe given by \citep{Wu2004,Wu2006}
in a future paper of this series.


\acknowledgments

We thank the anonymous referee for helpful comments which significantly improved the manuscript. This work is jointly supported by the 973 program under
grant 2012CB825601, the Chinese Academy of Sciences (KZZD-EW-01-4),
the National Natural Science Foundation of China (41204126, 41231068,
41274192, 41174151, 41031066, and 41074122), and the Specialized
Research Fund for State Key Laboratories. The work performed by STW is
supported by NSF-AGS1153323, and QH is supported by NSF SHINE AGS-1062050.
Numerical calculations were completed on our SIGMA Cluster computing system. Data from
observations are courtesy of NASA/{\SDO} and the HMI science teams.


\end{CJK*}
\end{document}